\documentclass[aps,prb,floatfix,twocolumn]{revtex4}%
\usepackage{graphicx}
\usepackage{amsfonts}
\usepackage{amssymb}
\usepackage{amsmath}
\usepackage{graphicx}
\usepackage{amsfonts}
\usepackage{amssymb}%
\providecommand{\U}[1]{\protect\rule{.1in}{.1in}}
\providecommand{\U}[1]{\protect\rule{.1in}{.1in}}
\providecommand{\U}[1]{\protect\rule{.1in}{.1in}}
\providecommand{\U}[1]{\protect\rule{.1in}{.1in}}
\providecommand{\U}[1]{\protect\rule{.1in}{.1in}}

\begin{document}
\title{Electron-photon coupling in Mesoscopic Quantum Electrodynamics}
\author{A. Cottet$^{1}$\thanks{To whom correspondence should be addressed :
cottet@lpa.ens.fr}, T. Kontos$^{1}$ and B. Dou\c{c}ot$^{2}$}
\affiliation{$^{1}${\normalsize Laboratoire Pierre Aigrain, Ecole Normale
Sup\'{e}rieure-PSL Research University, CNRS, Universit\'{e} Pierre et Marie
Curie-Sorbonne Universit\'{e}s, Universit\'{e} Paris Diderot-Sorbonne Paris
Cit\'{e}, 24 rue Lhomond, 75231 Paris Cedex 05, France}}
\affiliation{$^{2}${\normalsize Sorbonne Universit\'{e}s, Universit\'{e} Pierre et Marie
Curie, CNRS, LPTHE, UMR 7589, 4 place Jussieu, 75252 Paris Cedex 05}}
\date{\today}

\begin{abstract}
Understanding the interaction between cavity photons and electronic
nanocircuits is crucial for the development of Mesoscopic Quantum
Electrodynamics (QED). One has to combine ingredients from atomic Cavity QED,
like orbital degrees of freedom, with tunneling physics and strong cavity
field inhomogeneities, specific to superconducting circuit QED. It is
therefore necessary to introduce a formalism which bridges between these two
domains. We develop a general method based on a photonic pseudo-potential to
describe the electric coupling between electrons in a nanocircuit and cavity
photons. In this picture, photons can induce simultaneously orbital energy
shifts, tunneling, and local orbital transitions. We study in details the
elementary example of a single quantum dot with a single normal metal
reservoir, coupled to a cavity. Photon-induced tunneling terms lead to a
non-universal relation between the cavity frequency pull and the damping pull.
Our formalism can also be applied to multi quantum dot circuits, molecular
circuits, quantum point contacts, metallic tunnel junctions, and
superconducting nanostructures enclosing Andreev bound states or Majorana
bound states, for instance.

\end{abstract}
\maketitle
\author{}

\section{Introduction}

Cavity Quantum ElectroDynamics (Cavity QED) enables the study of the
interaction between light and matter at the most elementary level, thanks to
the achievement of a strong coupling between a single atom and a single photon
trapped in a microwave or optical cavity\cite{Raimond}. This paradigm has been
recently brought into superconducting circuits: artificial atoms consisting of
two level superconducting circuits have been coupled to superconducting
cavities\cite{Wallraff,Paik}, in the context of Circuit QED. These experiments
provide an ideal playground to test the basic laws of quantum mechanics
because they can be described in terms of simple models like the
Jaynes-Cummings Hamiltonian. However, such models conceal essential physical
differences between Cavity and Circuit QED. On the one hand, the coupling
between isolated atoms and cavity photons mainly occurs due to the sensitivity
of the atom electric dipole to the cavity electric field. This coupling
depends on microscopic details since the atomic dipole is set by the structure
of the atom electronic orbitals. Furthermore, one can generally perform the
"electric-dipole approximation" which assumes that the cavity field varies
little on the scale of the atomic system\cite{Cohen-Tannoudji}. On the other
hand, the behavior of submicronic superconducting circuits is essentially
insensitive to microscopic details due to the rigidity of the superconducting
phase\cite{circuitQED}. For instance, the behavior of a superconducting charge
qubit can be described with one macroscopic variable, i.e. the total charge of
a superconducting island\cite{Bouchiat,Nakamura}. This charge can vary due to
the Josephson coupling between the island and an external superconducting
reservoir. The coupling between the superconducting charge qubit and the
cavity is usually described in terms of a capacitive coupling between the
superconducting island and the cavity central conductor. As a result, the
chemical potential of the superconducting island is shifted proportionally to
the cavity electric potential\cite{Blais}. This picture implies strong
inhomogeneities of the photonic electric field on the scale of the
superconducting qubit, in contrast to what is generally considered in atomic
Cavity QED for a single atom.

Recent technological progress is enabling the development of a new type of
experiments where nanocircuits based on carbon nanotubes, semiconducting
nanowires, two-dimensional electron gases or graphene, are coupled to coplanar
microwave
cavities\cite{Delbecq1,Delbecq2,DQD1,Petersson,DQD2,DQD3,DQD4,DQD5,Schroer,Viennot,Liu,Liu2}%
. This paves the way for the development of "Mesoscopic QED", a denomination
introduced in a pioneering theory work\cite{Childress}. Mesoscopic QED opens
many possibilities because nanoconductors can be tunnel-coupled to various
types of fermionic reservoirs such as normal metals,
ferromagnets\cite{CottetSST} or superconductors\cite{Franceschi}, in a large
variety of geometries. So far, theoretical studies on Mesoscopic QED have
mainly focused on quantum dot
circuits\cite{Childress,SpinqubitsV,SqAE1,SqAE2,SqAE3,QubitJin,Lasing,LasingTheoPetta,DQDth,DT1,DT2,DT3,DT4,DT5,DT6,DT7,DT8,CPS1,CPS2,Hu,SingleDot}%
. Several configurations have been suggested to reach the strong coupling
regime between an electronic spin and cavity
photons\cite{SpinqubitsV,QubitJin,Hu,SqAE1,SqAE2,SqAE3}, or more generally, to
develop quantum computing
schemes\cite{QC1,QC2,QC3,QC4,QC5,QC6,QC7,Schmidt1,Schmidt2}. Mesoscopic QED
also tackles problems which go beyond the mechanics of closed two level
systems coupled to cavities, usually considered in Cavity or Circuit QED. The
interaction between electronic transport and the light-matter interaction
leads to a rich
phenomenology\cite{Childress,Lasing,LasingTheoPetta,DT1,DT2,DT3,DT4,DT5,DT6,DT7,DT8,CPS1,CPS2,Souquet,BergenfeldtSET}%
. Besides, coplanar cavities could be used as a powerful probe to reveal some
exotic properties of hybrid nanocircuits, like for instance the existence of
topological superconducting phases\cite{Trif}, Majorana quasiparticle
modes\cite{QC1,QC2,QC3,QC4,QC5,QC6,QC7,Schmidt1,Schmidt2,CottetMajos}, or
spin-entanglement in a Cooper pair beam splitter\cite{CPS1,CPS2}. On the
experimental side, pioneering works have focused on mesoscopic
rings\cite{Reulet} and metallic tunnel junctions\cite{Holst}. More recently,
experiments have been performed with single quantum
dots\cite{Delbecq1,Delbecq2} and double quantum dots (DQDs)
\cite{DQD1,Petersson,DQD2,DQD3,DQD4,DQD5,Schroer,Viennot,Liu,Liu2} with normal
metal reservoirs. Reaching the strong coupling regime between the charge
states of a DQD and a cavity is still a challenge\cite{Wallraff2}.
Nevertheless, interesting resonance phenomena have already been
observed\cite{DQD1,Petersson,DQD2,DQD3,DQD4,Viennot}. Several experiments have
also provided evidence for a modification of the cavity behavior by finite
bias transport through a DQD\cite{Viennot,Liu}, including a maser
effect\cite{Liu2}.

These recent developments call for a full description of the coupling between
a hybrid nanocircuit and cavity photons. One question naturally arises: is
Mesoscopic QED closer to atomic Cavity QED or superconducting Circuit QED?
What are the specificities of the coupling between a nanocircuit and a cavity?
So far, most theory works have considered a capacitive coupling between the
nanocircuit and the cavity central conductor, by analogy with Circuit
QED\cite{SingleDot,Lasing,DQDth,SpinqubitsV,DT1,DT2,DT3,DT4,DT5,DT6,DT7,DT8,CottetMajos,BergenfeldtSET}%
. This approach implies a coarse grained electric description of the
nanocircuit, and a concentration of the non-homogeneous photonic electric
field inside some capacitive elements. A few works have considered a direct
coupling between the motion of electrons trapped in the nanoconductors and the
bare cavity electric field, which is assumed to be constant on the scale of
the nanocircuit\cite{,SqAE1,SqAE2,SqAE3,Schmidt2,CPS1,Hu}. This is reminiscent
of the descriptions used in Cavity QED\cite{Cohen-Tannoudji}. In this article,
we introduce a description of Mesoscopic QED which bridges between these two
approaches. We use a model which focuses on conduction electrons tunneling
between the different elements of a nanocircuit. The tunneling electrons
occupy quasi-localized orbitals in each nanocircuit element, which recalls the
atomic orbital degree of freedom of Cavity QED. However, there also exists
collective plasmonic modes in the nanocircuit, which can screen at least
partially the cavity fields.\ We use a gauge-invariant Mesoscopic QED
Hamiltonian which accounts for the non-uniform screening of the cavity fields
inside the nanocircuit, and for the electromagnetic boundary conditions
provided by cavity conductors and voltage-biased nanocircuit DC gates. In the
limit where photon-induced magnetic effects are negligible, we can reexpress
the Mesoscopic QED Hamiltonian in terms of a scalar photonic pseudo-potential.
This picture unifies the different approaches used so far to describe
Mesoscopic QED devices, since the photonic pseudo-potential can vary linearly
with space in the case of a locally uniform photonic electric field (dipolar
coupling limit), as well as remain constant inside a given circuit element in
the limit of a coarse grained circuit model. In the framework of a tunneling
model, the photonic pseudo-potential leads to various types of linear
electron/photon coupling terms: cavity photons shift the orbital energies of
the different nanocircuit elements, but also induce simultaneously tunneling
and local orbital transitions. This general description can be used to study
the behavior of many different types of Mesoscopic QED devices. For instance,
it can be applied to quantum dot circuits, molecular circuits, quantum point
contacts, metallic tunnel junctions, and superconducting nanostructures
enclosing Andreev bound states or Majorana bound states. To illustrate the
richness of our approach, we study in details the elementary example of a
cavity coupled to a "quantum RC circuit", i.e. a single quantum dot coupled to
a single normal metal reservoir. The photon-induced tunneling terms between
the quantum dot and the reservoir induce a non-universal relation between the
cavity frequency pull and the cavity damping pull, contrarily to what is
expected with purely capacitive coupling schemes at low
temperatures\cite{RCbuttiker,RC1,RC2,RC3,RC4,RC5,RC6}.

This paper is organized as follows. In section II A, we discuss the
gauge-invariant Mesoscopic QED Hamiltonian, which involves a photonic vector
potential. In section II B, we perform a unitary transformation to obtain a
new Hamiltonian where the electron/photon coupling is due to the scalar
photonic pseudo-potential. In section II C, we reexpress the photonic
pseudo-potential scheme in the framework of a tunnelling model. In section II
D, we discuss the application of our formalism to the case of nanocircuits
with superconducting elements. In particular, we give an explicit Hamiltonian
for a nanostructure with Majorana bound states coupled to a cavity. In section
III, we work out in details the case of the quantum RC circuit coupled to a
cavity. Section IV concludes. Appendix A gives a detailed mathematical
justification for the form of the Hamiltonian of section II A, on the basis of
an effective model which separates physically the tunneling electrons
occupying individual orbital states in the different elements of a
nanocircuit, from the plasmonic collective modes of this nanocircuit. Appendix
B discusses the advantages of the photonic pseudo-potential scheme.

\section{General description of Mesoscopic QED}

\subsection{Gauge invariance and minimal coupling scheme}

A nanocircuit encloses a large variety of degrees of freedom. Our main
interest in this paper is the interaction of the cavity with the ensemble
$\mathcal{T}$ of the "tunneling" conduction electrons, which can occupy
orbital states in the different elements of a nanocircuit and tunnel between
them. However, there also exists plasmonic electronic modes which are
responsive for the screening of external fields from a massive metallic
electrode. Plasmons can also convey screening currents in the reservoir
electrical lines, which reequilibrate locally charge distributions on tunnel
junctions after a tunneling event. \begin{figure}[h]
\includegraphics[width=0.85\linewidth]{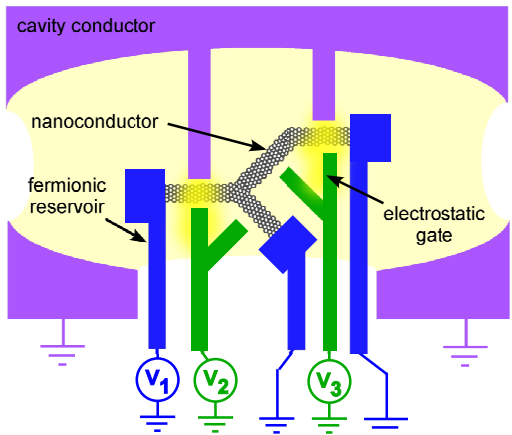}\caption{Scheme of a
loopless nanocircuit with source/drain electrodes (blue) and electrostatic
gates (green) embedded in a photonic cavity (purple). The yellow cloud
represents the photonic field. The cavity presents some protrusions (purple
stripes) fabricated to increase the photonic field inhomogeneities (darker
yellow areas).}%
\end{figure}The question of plasmonic modes in a nanocircuit is complicated
since nanoconductors have generally a low electronic density, which allows
only a partial screening of the cavity field\cite{Klinovaja}. In principle,
even the metallic contacts of the nanocircuit are not able to screen totally
the cavity field due to their thinness. In the limit where the nanocircuit
contacts are connected to outside electrical lines with a low impedance, it is
reasonable to assume that the dynamics of plasmonic modes in the nanocircuit
is very fast compared to the dynamics of the cavity modes and of the tunneling
electrons\cite{DCB}. In this case, one can assume that the plasmonic modes in
the nanocircuit generate a distribution of screening charges which is enslaved
to the position of the tunneling charges and to the value of the cavity field.
This produces a renormalization of the cavity field properties, and in
particular its spatial dependence nearby and inside the nanocircuit. It also
produces a renormalization of the Coulomb interactions between the tunneling
charges. In general, the Coulomb interactions between the nanocircuit charges
is also renormalized by the screening charges on the cavity conductors (see
Appendix A). The positive ions and valence electrons constituting the
crystalline structure of the nanocircuit can be treated in a mean field
approach, leading to an effective potential $V_{conf}(\overrightarrow{r})$
which confines tunneling electrons inside the nanocircuit. This confinement
potential can delimit quantum dots, or induce disorder effects in
nanoconductors, for instance. This approach is sufficient provided the
background charges producing $V_{conf}(\overrightarrow{r})$ and the microwave
cavity are far off resonant. In this framework, gauge invariance imposes the Hamiltonian%

\begin{equation}
\hat{H}_{tot}=\int d^{3}r\hat{\psi}^{\dagger}(\vec{r})\hat{h}_{\mathcal{T}%
}(\vec{r})\hat{\psi}(\vec{r})+\hat{H}_{Coul}+\hbar\omega_{0}\hat{a}^{\dag}%
\hat{a} \label{1}%
\end{equation}
with%
\begin{equation}
\hat{h}_{\mathcal{T}}(\vec{r})=\frac{1}{2m}\left(  \frac{\hbar\overrightarrow
{\nabla}_{\vec{r}}}{i}+e\overrightarrow{\hat{A}}(\vec{r})\right)  ^{2}%
-e\Phi_{harm}(\overrightarrow{r})-eV_{conf}(\overrightarrow{r})~\text{,}
\label{202}%
\end{equation}%
\begin{equation}
\hat{H}_{Coul}=\frac{e^{2}}{2}\int d^{3}rd^{3}r^{\prime}\hat{\psi}^{\dagger
}(\vec{r})\hat{\psi}^{\dagger}(\vec{r}^{\text{ }\prime})G(\vec{r},\vec
{r}^{\text{ }\prime})\hat{\psi}(\vec{r}^{\text{ }\prime})\hat{\psi}(\vec
{r})~\text{,} \label{coul}%
\end{equation}%
\begin{equation}
\overrightarrow{\hat{A}}(\vec{r})=\mathcal{\vec{A}}(\vec{r})i(\hat{a}-\hat
{a}^{\dag})~\text{,} \label{6}%
\end{equation}
and $e>0$ the elementary charge. The field operator $\hat{\psi}(\vec{r})$
includes all the tunneling charges of the nanocircuit. Above, $\hat
{h}_{\mathcal{T}}(\vec{r})$ is a single electron Hamiltonian, $G(\vec{r}%
,\vec{r}^{\text{ }\prime})$ describes the Coulomb interaction between the
tunneling electrons (see Appendix A for details), and $\hbar\omega_{0}\hat
{a}^{\dag}\hat{a}$ is the Hamiltonian of a (renormalized) cavity
electromagnetic mode. The potential $\Phi_{harm}(\overrightarrow{r})$ is due
to DC voltages applied on the nanocircuit electrostatic gates and to constant
charges on floating conductors like the cavity central conductor in a coplanar
geometry. The coupling between the cavity and the tunneling electrons arises
from the photonic vector potential term of Eq.(\ref{6}). For simplicity, we
have considered a single cavity mode with creation operator $\hat{a}^{\dag}$
and vector potential profile $\mathcal{\vec{A}}(\vec{r})$. This description
can be generalized straightforwardly to the multimode case by introducing a
sum on a cavity mode index in Eqs. (\ref{1}) and (\ref{6}). Note that
$\mathcal{\vec{A}}(\vec{r})$ can strongly vary on the scale of the nanocircuit.

The formal description of the electromagnetic field and the light/matter
interaction in Mesoscopic QED requires some care. In atomic cavity QED, the
cavity field is generally assumed to vary slowly on the scale of an
atom\cite{Cohen-Tannoudji}. In contrast, we have to take into account strong
spatial variations of the photonic field on the scale of a nanocircuit. In
particular, a protrusion of a cavity conductor can be used to increase the
photonic field locally, close to a given quantum dot
\cite{DQD1,DQD2,DQD3,Viennot} (See Fig. 1). The DC voltage-biased
electrostatic gates, used to control the nanocircuit spectrum, provide
supplementary boundary conditions on the electromagnetic field. Besides, we
have already mentioned that the plasmonic screening charges on the nanocircuit
conductors can modify the photonic vector potential profile, especially inside
and around fermionic reservoirs. In order to describe this complex reality and
justify mathematically the shape of Eqs.(\ref{1}-\ref{6}), we introduce in
Appendix A an effective model which separates physically the tunneling
electrons occupying individual orbital states in the nanocircuit from the
collective plasmonic modes of the nanocircuit. In this framework, one can
introduce a Hodge decomposition of the electromagnetic
field\cite{Hodge,Hodge2}, which can be obtained thanks to the use of an
auxiliary electrostatic Green's function. This leads to the terms in
$\overrightarrow{\hat{A}}(\vec{r})$, $G(\vec{r},\vec{r}^{\text{ }\prime})$ and
$\Phi_{harm}(\overrightarrow{r})$ in Eqs. (\ref{1}-\ref{6}).

In this section, we disregard the spin of charge carriers because we want to
focus on regimes where charge effects prevail. The introduction of the spin
degree of freedom in Hamiltonian (\ref{1}) would be straightforward by
invoking gauge invariance (see for instance section \ref{Ad}). The direct
coupling between the particles spin and the photonic magnetic field can
usually be disregarded because this coupling is expected to be very weak
unless collective excitations are considered in a large ensemble of
spins\cite{cri1,cri2,NoriReview}. From recent predictions,
real\cite{SqAE1,SqAE2,SqAE3} or effective\cite{SpinqubitsV,QubitJin,Hu}
spin-orbit coupling provide an alternative way to obtain a spin-photon coupling.

In Eq.(\ref{202}), the coupling of tunneling electrons to cavity photons
occurs through two terms, one in $\overrightarrow{\nabla}_{\vec{r}%
}.\overrightarrow{\hat{A}}+\overrightarrow{\hat{A}}.\overrightarrow{\nabla
}_{\vec{r}}$, corresponding to single photon transitions, and the other one in
$\hat{A}^{2}$, corresponding to two photon transitions. The second has no
reason to be disregarded in the general case, even if it has no structure in
the electronic sector. For instance, this term is crucial for determining the
existence of a Dicke quantum phase transition when many two-level systems are
coupled transversely to a cavity\cite{Nataf,Viehmann}. The $\hat{A}^{2}$ term
can also bring a non-negligible contribution to the cavity frequency pull
caused by one nanocircuit, which is a central quantity in Mesoscopic QED
experiments (see Appendix B). Hence, it is convenient to introduce a new
representation in which the $\hat{A}^{2}$ term is eliminated. This task is
completed in the next section.

\subsection{Photonic pseudo-potential scheme\label{pseudoP}}

In atomic Cavity QED, the Power-Zienau-Wooley (PZW) transformation formally
enables the elimination of a space-dependent $\hat{A}(\vec{r})^{2}$ term from
the system Hamiltonian\cite{Cohen-Tannoudji}. This generalizes the widely
known "electric-dipole" transformation used in the case where $\hat{A}(\vec
{r})$ can be considered as constant on the scale of an atom. However, the
natural variables of the PZW Hamiltonian are polarization and magnetization
densities associated to the charge distribution of an atomic system, which are
not directly useful in our case. To reexpress the PZW Hamiltonian in terms of
particles coordinates and momenta, it is necessary to perform a multipolar
development of the light/matter interaction\cite{Cohen-Tannoudji}. In
practice, this development is performed to a limited order, which means that
only a moderate space dependence of the photonic electric and magnetic fields
is taken into account. In this section, we perform a unitary transformation of
Hamiltonian (\ref{1}) in the same spirit as the PZW transformation, but with
specificities required for Mesoscopic QED. Upon disregarding magnetic effects,
we can take into account a strong space dependence of the photonic electric
field, thanks to the use of a scalar photonic pseudo-potential. This procedure
avoids the development of the light/matter interaction in terms of multipoles.
This is necessary to describe e.g. multi-quantum dot circuits with strongly
asymmetric capacitive couplings to the cavity. Furthermore, we use a
second-quantized description of charges, which is more convenient to describe
central phenomena of mesoscopic physics like tunneling but also
superconductivity (see next section).

To account for the possibly strong spatial variations of $\mathcal{\vec{A}%
}(\vec{r})$ on the scale of a nanocircuit, we define inside the volume of the
nanocircuit a reference point $\vec{r}_{0}$ and a continuous functional path
$C(\vec{r},\vec{r}^{~\prime})$ relating any points $\vec{r}$ and $\vec
{r}^{~\prime}$. We also define a photonic pseudo-potential $V_{\perp
}(\overrightarrow{r})(\hat{a}+\hat{a}^{\dag})$ with%
\begin{equation}
V_{\perp}(\overrightarrow{r})=\omega_{0}\int\nolimits_{C(\vec{r}_{0},\vec{r}%
)}\mathcal{\vec{A}}(\overrightarrow{r}^{\prime}).\overrightarrow{d\ell
}~\text{.}%
\end{equation}
Importantly, one can use $\vec{\nabla}_{\vec{r}}.V_{\perp}(\overrightarrow
{r})\simeq\omega_{0}\mathcal{\vec{A}}(\vec{r})$ inside the nanocircuit volume,
provided the flux of $\vec{\nabla}_{\vec{r}}\wedge\mathcal{\vec{A}}$ through
the nanocircuit can be disregarded. This criterion is generally satisfied,
considering the typical amplitude of the cavity magnetic field and the size of
nanoconductors. In this case, the arbitrariness on the choice of the contours
$C(\vec{r},\vec{r}^{~\prime})$ leads only to marginal effects. One exception
is nanocircuits with large loops leading to magnetic Aharonov-Bohm effects.
This case is beyond the scope of the rest of this article, which focuses
on\ photon-induced electric effects. We perform the transformation
\begin{equation}
\widetilde{H_{tot}}=\mathcal{U}^{\dag}H_{tot}\mathcal{U} \label{Unit}%
\end{equation}
with
\begin{equation}
\mathcal{U}=\exp\left(  \frac{e(\hat{a}-\hat{a}^{\dag})}{\hbar\omega_{0}}\int
d^{3}r~V_{\perp}(\overrightarrow{r})\hat{\psi}^{\dagger}(\vec{r})\hat{\psi
}(\vec{r})\right)  ~\text{.} \label{unit2}%
\end{equation}
This gives
\begin{align}
\widetilde{H_{tot}}  &  =\int d^{3}r\hat{\psi}^{\dagger}(r)\widetilde
{h}_{\mathcal{T}}(\vec{r})\hat{\psi}(\vec{r})+\hat{H}_{Coul}+\hbar\omega
_{0}\hat{a}^{\dag}\hat{a}\nonumber\\
&  +\hat{\mathcal{V}}(\hat{a}+\hat{a}^{\dag})+(\hat{\mathcal{V}}^{2}%
/\hbar\omega_{0}) \label{Htottild}%
\end{align}
with%
\begin{equation}
\hat{\mathcal{V}}=-e\int d^{3}rV_{\perp}(\overrightarrow{r})\hat{\psi
}^{\dagger}(\vec{r})\hat{\psi}(\vec{r})
\end{equation}
and $\widetilde{h}_{\mathcal{T}}(\vec{r})=-\hbar^{2}\Delta_{\overrightarrow
{r}}/2m-e\Phi_{harm}(\overrightarrow{r})-eV_{conf}(\overrightarrow{r})$. In
this new representation, the single electron potential and the Coulomb
interactions are renormalized by the term in $\hat{\mathcal{V}}^{2}$, but the
electron/photon coupling takes a simpler linear form. The motion of tunneling
electrons is modified by the pseudo-potential $V_{\perp}(\overrightarrow
{r})(\hat{a}+\hat{a}^{\dag})$. Note that the arbitrariness on the choice of
$\vec{r}_{0}$ is not an issue, since a change in $\vec{r}_{0}$ shifts
$V_{\perp}(\overrightarrow{r})$ by a global constant and leads to a unitary
transformation of $\widetilde{\hat{H}}_{tot}$. In the limit where the spatial
dependence of the photonic vector potential can be disregarded i.e. $\vec
{A}(\overrightarrow{r})\simeq\vec{A}_{0}$, one finds $V_{\perp}%
(\overrightarrow{r})=\vec{E}_{0}.\overrightarrow{r}(\hat{a}+\hat{a}^{\dag})$
with $\vec{E}_{0}(\overrightarrow{r})=\omega_{0}\vec{A}_{0}$. This corresponds
to the electric-dipole approximation frequently used in atomic Cavity QED. In
the opposite limit of a small enough conductor or a perfect screening of
$\vec{A}(\overrightarrow{r})$ inside this conductor, one can use a $V_{\perp
}(\overrightarrow{r})$ constant inside this conductor. This is reminiscent of
the capacitive network models frequently used to describe superconducting
charge qubits in Circuit QED. Note that we do not assume that the ensemble
$\mathcal{T}$ of the tunneling charges is neutral.

\subsection{Tunneling model and coarse graining of nanocircuits\label{tunnel}}

In this section, we consider a nanocircuit gathering quantum dots and
reservoirs connected by tunnel junctions. The effect of time-dependent
classical fields on tunneling through quantum dots has been studied since the
1990's\cite{Kouwenhoven,Platero}. Mesoscopic Circuit QED offers the
opportunity to study the interaction between microwave photons and quantum dot
circuits from a different perspective. Quantum dot circuits are often
described in terms of a tunneling model in which the dots and reservoirs
contain quasi-localized orbitals coupled by tunnel elements\cite{Schrieffer}.
This decomposition is instrumental in order to particularize the treatment of
the reservoirs and account for the irreversibility of transport processes or
other reservoir-induced damping effects. Importantly, the tunnel couplings
arise from an overlap between neighboring orbitals. This means that these
orbitals cannot be considered as orthogonal, except in the limit of weak
tunnel couplings (see for instance \cite{Fransson}). However, in practice, it
may be necessary to take into account a strong inter-dot tunneling. For
example, in the case of a DQD, there can exist a strong splitting between
bonding and antibonding states, which can be resonant with the
cavity\cite{DQD1,Petersson,DQD2,DQD3,DQD4,Viennot}. Depending on the absolute
energy scale of the DQD confinement potential, it might be necessary to go to
the limit of a non-perturbative inter-dot tunnel coupling to obtain such a
splitting. In order to circumvent the difficulty related to the
non-orthogonality of the nanocircuit orbitals, we divide the nanocircuit into
the ensemble $\mathcal{D}$ of tunnel-coupled quantum dots from one side, and
the different reservoirs with an index $R_{p}$ with $p\in\{1,...,N\}$, from
the other side. We use an index $o\in\{\mathcal{D},R_{1},$...$,R_{N}\}$ to
denote these different elements, and an index $j$ to denote the different
orbitals in a given element $o$. We assume that the low energy spectrum for
the whole ensemble $\mathcal{D}$ of the quantum dots can be determined in the
absence of the reservoirs as the discrete spectrum of a potential profile with
one or several potential wells. This procedure can be performed for an
arbitrary inter-dot tunneling strength and leads to exactly orthogonal
orbitals, so that $\{\hat{c}_{\mathcal{D}j}^{\dag},\hat{c}_{\mathcal{D}%
j^{\prime}}\}=\delta_{j,j^{\prime}}$. One can also define exactly orthogonal
orbitals in the isolated reservoir $R_{p}$, with $p\in\{1,...,N\}$, so that
$\{\hat{c}_{R_{pj}}^{\dag},\hat{c}_{R_{p^{\prime}j^{\prime}}}\}=\delta
_{p,p^{\prime}}\delta_{j,j^{\prime}}$. In a second step, one can evaluate the
tunnel coupling $t_{\mathcal{D}j,R_{p}j^{\prime}}$ between the orbitals $j$
and $j^{\prime}$ of $\mathcal{D}$ and $R_{p}$ from the overlap between their
wavefunctions. This gives $\{\hat{c}_{\mathcal{D}j}^{\dag},\hat{c}%
_{R_{p}j^{\prime}}\}=0$ only at lowest order in tunneling\cite{Prange}.
However, using a weak tunneling between $\mathcal{D}$ and $R_{p}$ is not a
severe restriction since the large density of states in the reservoirs can
compensate for the smallness of the tunnel coupling elements, and lead to a
large tunneling rate compatible with the Kondo effect, for instance. To
summarize, in the absence of the cavity, the tunnel Hamiltonian writes%
\begin{equation}
\hat{H}_{e}=\sum_{o,j}\varepsilon_{oj}\hat{c}_{oj}^{\dag}\hat{c}_{oj}%
+\sum_{o\neq o^{\prime},j,j^{\prime}}(t_{oj,o^{\prime}j^{\prime}}\hat
{c}_{o^{\prime}j^{\prime}}^{\dag}\hat{c}_{oj}+H.c.) \label{htun}%
\end{equation}
where the creation operator $\hat{c}_{oj}^{\dag}$ corresponds to an orbital
$j$ with energy $\varepsilon_{oj}$ and wavefunction $\varphi_{oj}(\vec{r})$
mainly localized inside element $o\in\{\mathcal{D},R_{1},$...$,R_{N}\}$. One
can use $\{\hat{c}_{oj}^{\dag},\hat{c}_{o^{\prime}j^{\prime}}\}=\delta
_{o,o^{\prime}}\delta_{j,j^{\prime}}$ at lowest order in the dot/reservoir
tunnel couplings.

We now reexpress the photonic pseudo-potential scheme of the previous section
in the framework of the tunneling model. For this purpose, one needs to
decompose the field operator $\hat{\psi}^{\dagger}(\vec{r})$ associated to
tunneling charges on the nanocircuit orbital states. At lowest order in the
dot/reservoir tunneling, one can use\cite{Prange}
\begin{equation}
\hat{\psi}^{\dagger}(\vec{r})=\sum_{o,j}\varphi_{oj}^{\ast}(\vec{r})\hat
{c}_{oj}^{\dag}~\text{.} \label{dec}%
\end{equation}
Hence, Hamiltonian (\ref{Htottild}) directly gives%
\begin{equation}
\hat{H}_{tot}^{tun}=\hat{H}_{e}+\hat{H}_{Coul}^{tun}+\hat{h}_{int}(\hat
{a}+\hat{a}^{\dag})+\hbar\omega_{0}\hat{a}^{\dag}\hat{a} \label{Ht}%
\end{equation}
with%
\begin{equation}
\hat{h}_{int}=\sum_{o,j}\alpha_{oj}\hat{c}_{oj}^{\dag}\hat{c}_{oj}%
+\sum_{oj\neq o^{\prime}j^{\prime}}(\gamma_{oj,o^{\prime}j^{\prime}}\hat
{c}_{o^{\prime}j^{\prime}}^{\dag}\hat{c}_{oj}+H.c.)~\text{.}%
\end{equation}
Note that in principle, the term in $\hat{\mathcal{V}}^{2}$ from
Eq.(\ref{Htottild}) renormalizes $\varepsilon_{oj}$, $t_{oj,o^{\prime
}j^{\prime}}$ and the Coulomb interaction term $\hat{H}_{Coul}^{tun}$, but
this is not essential for the physics we discuss below.

Since the photonic pseudo-potential modifies the potential landscape seen by
the tunneling charges, it can naturally affect all the parameters in the
tunneling model. First, cavity photons shift the orbital energy $\varepsilon
_{oj}$, with a coupling coefficient%
\begin{equation}
\alpha_{oj}=-e%
%TCIMACRO{\tint }%
%BeginExpansion
{\textstyle\int}
%EndExpansion
dr^{3}\left\vert \varphi_{oj}(\vec{r})\right\vert ^{2}V_{\perp}(\vec
{r})~\text{.} \label{jj1}%
\end{equation}
In general, $\alpha_{oj}$ strongly depends on the indexes $o$ and $j$ due to
the space dependence of $V_{\perp}(\vec{r})$. This makes the cavity-induced
orbital energy shifts particularly relevant experimentally. For a standard
metallic reservoir, it is reasonable to disregard the dependence of
$\alpha_{oj}$ on the orbital index $j$, i.e. $\alpha_{oj}\simeq\alpha_{o}$,
because the properties of the electronic wavefunctions in the reservoir can be
considered as constant near the Fermi energy\cite{reservoir}. In this
framework, a behavior similar to the capacitive coupling of the cavity central
conductor to the reservoirs is recovered. This type of reservoir/cavity
coupling enables one to interpret data obtained for a quantum dot coupled to
two normal metal leads\cite{Delbecq1,Delbecq2}. In contrast, $\alpha_{oj}$ may
strongly depend on $j$ for $o=\mathcal{D}$, as illustrated for instance by the
example of a DQD with an asymmetric $V_{\perp}(\vec{r})$%
\cite{DQD1,DQD2,DQD3,DQD4,Schroer,Viennot,Liu}.

The Hamiltonian $\hat{H}_{orb}$ also contains coupling terms in
\begin{equation}
\gamma_{oj,o^{\prime}j^{\prime}}=-e%
%TCIMACRO{\tint }%
%BeginExpansion
{\textstyle\int}
%EndExpansion
dr^{3}\varphi_{oj}^{\ast}(\vec{r})\varphi_{o^{\prime}j^{\prime}}(\vec
{r})V_{\perp}(\vec{r}) \label{jj2}%
\end{equation}
with $oj\neq o^{\prime}j^{\prime}$. These terms include photon-induced
tunneling terms in $\gamma_{oj,o^{\prime}j^{\prime}}$, with $o\neq o^{\prime}%
$, and photon-induced transitions internal to $o$, in $\gamma_{oj,oj^{\prime}%
}$. In principle, all the above cavity/nanocircuit coupling terms can coexist.
In practice, we expect the terms in $\alpha_{oj}$ to be dominant over the
tunnel terms in $\gamma_{oj,o^{\prime}j^{\prime}}$ with $o\neq o^{\prime}$,
which involve a weak overlap between wavefunctions. Nevertheless, as shown in
the next section, the signatures of the photon-induced tunneling terms can be
boosted by the large number of states they affect in the reservoirs. If
$V_{\perp}(\vec{r})$ varies slowly inside $\mathcal{D}$, the photon-induced
transitions internal to $\mathcal{D}$ are negligible, i.e. $\gamma
_{\mathcal{D}j,\mathcal{D}j^{\prime}}\simeq0$, since $\left\vert
\varphi_{\mathcal{D}j}\right\rangle $ and $\left\vert \varphi_{\mathcal{D}%
j^{\prime}}\right\rangle $ are orthogonal by definition. However, these
transitions become possible if the cavity field is weakly screened ($V_{\perp
}(\overrightarrow{r})\simeq\vec{E}_{0}.\overrightarrow{r}(\hat{a}+\hat
{a}^{\dag})$), or if a protrusion of a cavity conductor is placed close to one
of the quantum dots in order to reinforce the spatial variations of $V_{\perp
}(\vec{r})$\cite{DQD1,DQD2,DQD3,Viennot}. Transitions in $\gamma
_{\mathcal{D}j,\mathcal{D}j^{\prime}}$ can be particularly interesting in case
of real or artificial spin orbit coupling which mixes spin and orbital states.
In this case, the $\gamma_{\mathcal{D}j,\mathcal{D}j^{\prime}}$ terms can
correspond to spin-transitions inside a single quantum
dot\cite{SqAE1,SqAE2,SqAE3} or a DQD\cite{SpinqubitsV,QubitJin,Hu}. In the
reservoirs, the $\gamma_{R_{p}j,R_{p}j^{\prime}}$ terms are probably always
non-resonant due to strong energy relaxation. Hence, we will assume that these
terms do not need to be treated explicitly because they can be included in a
renormalization of the states $\left\vert \varphi_{R_{p}j}\right\rangle $.

In practice, the potential $V_{\perp}(\vec{r})$ can be evaluated numerically
by removing all nanoconductors from the device and simulating the microwave
electromagnetic field in the cavity in the presence of the metallic gates,
sources and drains of the nanocircuit. This can be done using standard
microwave simulation tools, which can take into account imperfect metals. It
enables a realistic evaluation of the elements $\alpha_{oj}$ and
$\gamma_{oj,o^{\prime}j^{\prime}}$. Even if $V_{\perp}(\vec{r})$ is not
calculated numerically but replaced by a phenomenological expression, the
expressions (\ref{jj1}) and (\ref{jj2}) remain interesting because they set
constraints between the different $\alpha_{oj}$ and $\gamma_{oj,o^{\prime
}j^{\prime}}$, which all depend on the same $V_{\perp}(\vec{r})$ profile.

We do not give details on the Coulomb interaction term $\hat{H}_{Coul}^{tun}$
which stems directly from Eq.(\ref{Htottild}). If Coulomb blockade in the
nanocircuit is already strong in the absence of the cavity, the effect of the
cavity on interactions may be disregarded or treated perturbatively. In the
next section, we discuss another situation where we assume that the effects of
$\hat{H}_{Coul}^{tun}$ are negligible due to the large tunnel rate between a
quantum dot and a normal metal reservoir. In the general case, Coulomb
interactions between the tunneling charges can lead to a large variety of
effects, which we will not discuss in this work.

To conclude, we obtain a decomposition of the system Hamiltonian in terms of
nanocircuit elements connected by tunnel couplings. While this is reminiscent
of the coarse graining description of superconducting microcircuits frequently
used in Circuit QED, a full analogy is not possible due to the presence of the
orbital degree of freedom in the Mesoscopic QED case. Due to this orbital
degree of freedom, we obtain a large variety of electron/photon coupling
terms. In this section, we have discussed the case of quantum dot circuits
which raises most experimental efforts so far. However, the above approach can
be used for any other type of system in which a decomposition in terms of a
tunneling model is relevant. This includes quantum point contacts, molecular
circuits, metallic tunnel junctions, and hybrid superconducting systems
enclosing Andreev bound states or Majorana bound states, for instance. This
last case is discussed in more details in section \ref{supra}.

\subsection{Discussion on the theoretical context}

Many theoretical works on Mesoscopic QED use as a starting basis a tunnel
Hamiltonian where all parameters (orbital energies, tunnel rates,...) are
perturbed by corrections proportional to the cavity electric field. These
empirical approaches lack of a formal justification and can lead to unphysical
predictions. Our work provides a rigorous framework for the use of tunnelling
models with a linear light/matter interaction (the superconducting case will
be discussed explicitly in section \ref{supra}).

As already mentioned above, our approach proceeds along a spirit similar to
the PZW transformation, but with modifications necessary to take into account
the specificities of Mesoscopic QED, among which a strong space dependence of
the cavity electric field on the scale of a nanocircuit, and boundary
conditions provided by cavity conductors and voltage-biased DC electrostatic
gates. Interestingly, the effect of boundary conditions provided by grounded
conductors on the PZW transformation has been discussed recently\cite{Vukics}.
However, this Ref. considers only neutral atomic systems affected by locally
constant cavity fields (dipolar coupling limit).

\subsection{Mesoscopic QED with superconducting nanocircuits\label{supra}}

\subsubsection{Minimal coupling scheme with superconducting
nanocircuits\label{Ad}}

In the above sections, superconductivity was not explicitly taken into
account. It is important to discuss the generalization of our approach to the
case of superconducting hybrid nanocircuits, considering the present interest
for Majorana fermions\cite{Mourik} or Andreev bound states for
instance\cite{Bretheau}. For simplicity, we consider below standard BCS
superconducting correlations characterized in the mean field approach by a
superconducting gap $\Delta(\vec{r})$ coupling electrons and holes with
opposite spins. The gauge invariant minimal coupling Hamiltonian of
Eq.(\ref{1}) can be generalized as%
\begin{equation}
\hat{H}_{tot}^{S}=\hat{H}_{tot}+\int d^{3}r\left(  \Delta(\vec{r}%
)e^{2\hat{\Phi}(\vec{r})}\hat{\psi}_{\uparrow}^{\dagger}(\vec{r})\hat{\psi
}_{\downarrow}^{\dagger}(\vec{r})+H.c.\right)  \label{Hsupr}%
\end{equation}
with $\hat{\psi}_{\sigma}^{\dagger}(\vec{r})$ the field operator for tunneling
electrons with spin $\sigma\in\{\uparrow,\downarrow\}$ and%
\begin{equation}
\hat{\Phi}(\vec{r})=e(\hat{a}-\hat{a}^{\dag})V_{\perp}(\overrightarrow
{r})/\hbar\omega_{0}~\text{.}%
\end{equation}
The term $\hat{H}_{tot}$ is the straightforward generalization of Eq.(\ref{1})
to the spin dependent case, i.e. the single electron term $\hat{h}%
_{\mathcal{T}}(\vec{r})$ can have a structure in spin-space, due to magnetic
fields and/or spin-orbit coupling, for instance. Hence, $\hat{\psi}(\vec{r})$
must be replaced by $^{t}[\hat{\psi}_{\uparrow}(\vec{r}),\hat{\psi
}_{\downarrow}(\vec{r})]$ in the first term of Eq.(\ref{1}) and by $\hat{\psi
}_{\uparrow}(\vec{r})+\hat{\psi}_{\downarrow}(\vec{r})$ in Eq.(\ref{coul}).
Due to gauge invariance, $\hat{h}_{\mathcal{T}}(\vec{r})$ can involve the
electron momentum through $-i\hbar\overrightarrow{\nabla}_{\vec{r}%
}+e\overrightarrow{\hat{A}}(\vec{r})$ only. Importantly, we define
$\Delta(\vec{r})$ as a gauge invariant quantity, so that the phase parameter
$\hat{\Phi}$ ensures the gauge invariance the pairing term of $\hat{H}%
_{tot}^{S}$. The form used above for $\hat{\Phi}$ is valid provided
photon-induced magnetic effects can be disregarded, in agreement with the
approach of section \ref{pseudoP}.

\subsubsection{Photonic pseudo-potential scheme with superconducting
nanocircuits\label{PPP}}

Using Eqs.(\ref{Unit}) and (\ref{unit2}) with $\hat{\psi}^{\dagger}(\vec
{r})\hat{\psi}(\vec{r})$ replaced by $\Sigma_{\sigma}\hat{\psi}_{\sigma
}^{\dagger}(\vec{r})_{\sigma}\hat{\psi}(\vec{r})$, Hamiltonian (\ref{Hsupr})
is transformed into%
\begin{equation}
\widetilde{\hat{H}_{tot}^{S}}=\widetilde{\hat{H}_{tot}}+\int d^{3}r\left(
\Delta(\vec{r})\hat{\psi}_{\uparrow}^{\dagger}(\vec{r})\hat{\psi}_{\downarrow
}^{\dag}(\vec{r})+H.c.\right)  \label{bim}%
\end{equation}
where $\widetilde{\hat{H}_{tot}}$ is again the generalization of
Eq.(\ref{Htottild}) to the spin dependent case. Importantly, Eq.(\ref{bim})
shows that in the photonic pseudo-potential scheme, the phase of the gap term
is not affected by photons anymore. This represents one more advantage of this
scheme. This result remains valid in the case of d-wave or p-wave
superconducting correlations\cite{geneBCS}.

\subsubsection{Mesoscopic QED with Bogoliubov-De Gennes
equations\label{BdGsec}}

In the case where electron-electron interactions can be disregarded,
Bogoliubov-De Gennes equations are a widely used approach\cite{BdG}, which
enables a diagonalization of the circuit Hamiltonian in terms of
quasiparticles with creation operator $c_{n}^{\dagger}$. One can define
quasiparticles modes in the absence of the cavity, by using%
\begin{equation}
c_{n}^{\dagger}=%
%TCIMACRO{\tint }%
%BeginExpansion
{\textstyle\int}
%EndExpansion
d^{3}r~[\hat{\psi}_{\uparrow}^{\dag}(\overrightarrow{r}),\hat{\psi
}_{\downarrow}^{\dag}(\overrightarrow{r}),\hat{\psi}_{\uparrow}%
(\overrightarrow{r}),\hat{\psi}_{\downarrow}(\overrightarrow{r})].w_{n}%
(\overrightarrow{r})
\end{equation}
with a spinorial wavefunction
\[
w_{n}(z)=^{t}[u_{\uparrow}^{n}(z),u_{\downarrow}^{n}(z),v_{\downarrow}%
^{n}(z),-v_{\uparrow}^{n}(z)]
\]
fulfilling
\begin{equation}
h_{eff}(\vec{r})x_{n}(\vec{r})=E_{n}x_{n}(\vec{r})
\end{equation}
with $x_{n}(z)=^{t}[u_{\uparrow}^{n}(z),u_{\downarrow}^{n}(z),-v_{\uparrow
}^{n}(z),-v_{\downarrow}^{n}(z)]$ and
\begin{equation}
h_{eff}(\vec{r})=\left[
\begin{array}
[c]{cc}%
\widetilde{h}_{\mathcal{T}}(\vec{r}) & \Delta(\vec{r})\sigma_{0}\\
\Delta^{\ast}(\vec{r})\sigma_{0} & -\sigma_{y}\widetilde{h}_{\mathcal{T}%
}^{\ast}(\vec{r})\sigma_{y}%
\end{array}
\right]  ~\text{.}%
\end{equation}
Above, $\sigma_{0}$ and $\sigma_{y}$ are the identity and the second Pauli
matrices in spin space. From the definition of $c_{n}^{\dagger}$, one gets
\begin{align}
\widetilde{\hat{H}_{tot}^{S}} &  =\sum_{n}E_{n}c_{n}^{\dagger}c_{n}%
+\hbar\omega_{0}\hat{a}^{\dag}\hat{a}\label{BdG}\\
&  +\sum_{n,n^{\prime}}\left(  \mathcal{D}_{nn^{\prime}}^{(1)}c_{n}^{\dagger
}c_{n^{\prime}}+\mathcal{D}_{nn^{\prime}}^{(2)}\right)  (\hat{a}+\hat{a}%
^{\dag})\nonumber\\
&  +\sum_{n,n^{\prime}}\left(  \mathcal{D}_{nn^{\prime}}^{(3)}c_{n}^{\dagger
}c_{n^{\prime}}^{\dagger}+H.c.\right)  (\hat{a}+\hat{a}^{\dag})\nonumber
\end{align}
with%
\begin{equation}
\mathcal{D}_{nn^{\prime}}^{(1)}=-e%
%TCIMACRO{\tint }%
%BeginExpansion
{\textstyle\int}
%EndExpansion
d^{3}r~w_{n}^{\dag}(\overrightarrow{r})V_{\perp}(\overrightarrow{r})\tau
_{z}w_{n^{\prime}}(\overrightarrow{r})~\text{,}\label{Htot}%
\end{equation}%
\begin{equation}
\mathcal{D}_{nn^{\prime}}^{(2)}=-e%
%TCIMACRO{\tint }%
%BeginExpansion
{\textstyle\int}
%EndExpansion
d^{3}r~V_{\perp}(\overrightarrow{r})~(\left\vert v_{\downarrow}^{n}%
(z)\right\vert ^{2}+\left\vert v_{\uparrow}^{n}(z)\right\vert ^{2})~\text{,}%
\end{equation}%
\begin{equation}
\mathcal{D}_{nn^{\prime}}^{(3)}=-e%
%TCIMACRO{\tint }%
%BeginExpansion
{\textstyle\int}
%EndExpansion
d^{3}r~w_{n}^{\dag}(\overrightarrow{r})V_{\perp}(\overrightarrow{r})(\tau
_{x}+i\tau_{y})w_{n^{\prime}}^{\ast}(\overrightarrow{r})/2\label{ho}%
\end{equation}
and $\tau_{x}$, $\tau_{y}$ , $\tau_{z}$ Pauli matrices in Nambu space.

\subsubsection{Mesoscopic QED with Majorana fermions}

In this section, we discuss the case of a nanocircuit enclosing Majorana bound
states, coupled to a cavity\textbf{.} In principle, these self-adjoint bound
states can appear at the interface between the topological and non-topological
superconducting phases of a nanostructure\cite{Leijnse,Alicea,Beenakker}.
Although an indirect Majorana bound state/cavity coupling mediated by a
superconducting quantum bit or a Josephson junction has been considered in
many works\cite{QC1,QC2,QC3,QC4,QC5,QC6,QC7}, a direct coupling can also
exist\cite{CottetMajos}. Using our formalism, we can provide a general
Hamiltonian describing such a situation. The direct Majorana bound
state/cavity coupling occurs due an overlap between neighboring Majorana bound
states, caused by the finite size of the nanocircuit. This overlap is
naturally dependent on the photonic pseudo-potential. Proposals to obtain
Majorana bound states in condensed matter systems rely on the use of
superconducting elements, which calls for the use of the results of section
\ref{supra}. For simplicity, we disregard Coulomb interactions and use the
framework of the Bogoliubov-De Gennes equations of section \ref{BdGsec}. To
remain general, we do not specify the details of the system. We consider an
ensemble of Majorana bound states with creation operators $C_{M,n}^{\dagger
}=C_{M,n}$ and wavefunction $\varphi_{M,n}(\overrightarrow{r})$, localized
inside a nanoconductor. In general, the nanoconductor also encloses a
continuum of ordinary fermionic states $C_{S,n}^{\dagger}$ with wavefunction
$\varphi_{S,n}(\overrightarrow{r})$ located above an energy gap. We assume
that the nanoconductor is tunnel coupled to a fermionic reservoir with orbital
states $C_{R,n}^{\dagger}$ with wavefunctions $\varphi_{R,n}(\overrightarrow
{r})$. Such a reservoir can be used to measure the conductance through the
Majorana system\cite{Mourik}. In the limit of weak tunneling, one can
reexpress Eq.(\ref{BdG}) by using the ensemble of the operators $C_{o,n}%
^{\dag}$ with $o\in\{M,R,S\}$, following similar considerations as for the
tunneling model of section \ref{tunnel}. This gives:%
\begin{align}
\widetilde{\hat{H}_{tot}^{S}}  &  =\sum_{n<n^{\prime}}2i\theta_{n,n^{\prime}%
}C_{M,n}C_{M,n^{\prime}}+\sum_{o\in\{R,S\},n}\varepsilon_{n}^{o}C_{o,n}^{\dag
}C_{o,n}\nonumber\\
&  +\sum_{n,n^{\prime}}C_{M,n}(t_{n,n^{\prime}}^{MR}C_{R,n^{\prime}}^{\dag
}-\left(  t_{n,n^{\prime}}^{MR}\right)  ^{\ast}C_{R,n^{\prime}})+\hbar
\omega_{0}\hat{a}^{\dag}\hat{a}\nonumber\\
&  +\sum_{\substack{o,o^{\prime}\in\{M,S,R\}\\n,n^{\prime}}}\left(
\gamma_{on,o^{\prime}n^{\prime}}^{(1)}C_{o,n}^{\dag}C_{o,n^{\prime}}%
+\gamma_{on,o^{\prime}n^{\prime}}^{(2)}\right)  (\hat{a}^{\dag}+\hat
{a})\nonumber\\
&  +\sum_{\substack{o,o^{\prime}\in\{M,S,R\}\\n,n^{\prime}}}\left(
\gamma_{on,o^{\prime}n^{\prime}}^{(3)}C_{o,n}^{\dag}C_{o,n^{\prime}}^{\dag
}+H.c.\right)  (\hat{a}^{\dag}+\hat{a})~\text{.} \label{tt}%
\end{align}
Above, the coefficients $\gamma_{on,o^{\prime}n^{\prime}}^{(i)}$ have
expressions similar to the $\mathcal{D}_{nn^{\prime}}^{(i)}$ coefficients of
Eqs. (\ref{Htot}-\ref{ho}) with $w_{n(n^{\prime})}(\overrightarrow{r})$
replaced by $\varphi_{o(o^{\prime}),n(n^{\prime})}(\overrightarrow{r})$. We
use energies $\varepsilon_{n}^{o}$ for the orbitals in the reservoir ($o=R$)
and the nanoconductor continuum of states ($o=S$), and a tunnel coupling
parameter $\theta_{n,n^{\prime}}$ between overlapping Majorana bound states.
The Majorana bound states and the reservoir are connected by tunnel coupling
terms in $t_{n,n^{\prime}}^{MR}$. For simplicity, we have omitted the DC
tunnel coupling between the reservoir and the nanoconductor continuum of
states. Photons can shift all the above mentioned parameters, due to the two
last lines of Eq.(\ref{tt}). In particular, photons can modify the tunnel
coupling between neighboring Majorana bound states, due to the terms in
$\gamma_{Mn,Mn^{\prime}}^{(i)}$, as discussed in Ref.\cite{CottetMajos}. Note
that in Hamiltonian (\ref{tt}), there is no DC coupling between the Majorana
bound states and the nanoconductor continuum of states, which are assumed to
be orthogonal by construction of $C_{M,n}$ and $C_{S,n^{\prime}}$. However,
there can be photon-induced transitions between the Majorana bound states and
the nanocircuit continuum of states due to terms in $\gamma_{Mn,Sn^{\prime}%
}^{(i)}$ and $\gamma_{Sn^{\prime},Mn}^{(i)}$. Effects related to these last
transitions have been discussed in \cite{Trif}. Our approach provides a
general justification for such terms.

\section{Example: Quantum RC circuit in a cavity}

\subsection{Hamiltonian of the system}

We now give an example of a specific prediction given by Hamiltonian
(\ref{Ht}). We consider a cavity coupled to a "quantum RC circuit" made out of
a single quantum dot tunnel contacted to a single normal metal reservoir. We
assume that the dot has a large intrinsic level spacing so that a single dot
orbital with energy $\varepsilon_{d}$ needs to be considered. The dot is
capacitively coupled to a DC gate electrode, which enables one to tune
$\varepsilon_{d}$. We study the effects of the dot on the photonic mode with
frequency $\omega_{0}/2\pi$. Following the previous section, the total
Hamiltonian of the system is:%
\begin{equation}
\hat{H}_{RC+cav}=\hat{H}_{e}+\hat{h}_{int}(\hat{a}+\hat{a}^{\dag})+\hbar
\omega_{0}\hat{a}^{\dag}\hat{a}+\hat{H}_{bath}+\hat{H}_{RF}~\text{.}%
\end{equation}
The Hamiltonian
\begin{equation}
\hat{H}_{e}=\varepsilon_{d}\hat{c}_{d}^{\dag}\hat{c}_{d}+%
%TCIMACRO{\tint }%
%BeginExpansion
{\textstyle\int}
%EndExpansion
d\varepsilon N_{0}\nu(\varepsilon)(\varepsilon\hat{c}^{\dag}(\varepsilon
)\hat{c}(\varepsilon)+t\hat{c}_{d}^{\dag}\hat{c}(\varepsilon)+t^{\ast}\hat
{c}^{\dag}(\varepsilon)\hat{c}_{d})
\end{equation}
of the dot and its reservoir involves the lead density of states $N_{0}%
\nu(\varepsilon)$. For later use we have introduced a cutoff $\nu
(\varepsilon)=D^{2}/(D^{2}+\varepsilon^{2})$. The terms in $t$ describes
tunneling between the dot and the reservoir, whose orbitals correspond to
fermionic creation operators $\hat{c}_{d}^{\dag}$ and $c^{\dag}(\varepsilon)$,
respectively, with $\{\hat{c}_{d},\hat{c}_{d}^{\dag}\}=1$ and $\{\hat
{c}(\varepsilon),\hat{c}^{\dag}(\varepsilon^{\prime})\}=\delta(\varepsilon
-\varepsilon^{\prime})/N_{0}\nu(\varepsilon)$. One can use extra bosonic modes
$\hat{b}^{\dag}(\epsilon)$ with $[\hat{b}(\epsilon),\hat{b}^{\dag}%
(\epsilon^{\prime})]=\delta(\epsilon-\epsilon^{\prime})/n_{0}$ to describe the
intrinsic damping of the cavity, i.e.%
\begin{equation}
\hat{H}_{bath}=%
%TCIMACRO{\tint }%
%BeginExpansion
{\textstyle\int}
%EndExpansion
d\epsilon n_{0}(\epsilon\hat{b}^{\dag}(\epsilon)\hat{b}(\epsilon)+\tau\hat
{b}^{\dag}(\epsilon)\hat{a}+\tau^{\ast}\hat{a}^{\dag}\hat{b}(\epsilon
))~\text{.}%
\end{equation}
The coupling between the quantum dot circuit and the cavity is described by
the term $\hat{h}_{int}(\hat{a}+\hat{a}^{\dag})$ with%
\begin{equation}
\hat{h}_{int}=\alpha\hat{c}_{d}^{\dag}\hat{c}_{d}+%
%TCIMACRO{\tint }%
%BeginExpansion
{\textstyle\int}
%EndExpansion
d\varepsilon N_{0}\nu(\varepsilon)(\gamma\hat{c}_{d}^{\dag}\hat{c}%
(\varepsilon)+\gamma^{\ast}\hat{c}^{\dag}(\varepsilon)\hat{c}_{d})~\text{.}%
\end{equation}

It is not necessary to write explicitly the coupling between the chemical
potential of the reservoir and the cavity because since there is a single
reservoir, it is only the difference between the potentials of the dot and the
reservoir which matters. Note that the terminology "quantum RC circuit" mainly
refers to the fact that no DC current can flow through the device and
dissipation is due to transport through a single tunnel barrier. However, in
the general case, the concept of a local capacitor is not sufficient to
describe the coupling of the electromagnetic field to the device, since the
photonic pseudo potential modifies $\varepsilon_{d}$ and $t$ simultaneously.
In practice, one can measure the response of the cavity to a microwave drive
with frequency $\omega_{RF}/2\pi$, which can be described by%
\begin{equation}
\hat{H}_{RF}=\varepsilon_{RF}e^{i\omega_{RF}t}\hat{a}+\varepsilon_{RF}^{\ast
}e^{-i\omega_{RF}t}\hat{a}^{\dag}~\text{.} \label{coupl}%
\end{equation}
Above, we have disregarded Coulomb interactions between the tunneling charges
(including the contribution from the $\hat{\mathcal{V}}^{2}/\hbar\omega_{0}$
term arising from the unitary transformation of section II.B), because we
assume that these interactions are smaller than the tunnel rate $\Gamma
=\pi\left\vert t\right\vert ^{2}N_{0}$ between the dot and the reservoir. Such
a simplification is frequently performed for interpreting standard current
measurements in open quantum dot circuits\cite{Gabelli,Liang}. In practice,
one can check the relevance of this hypothesis by measuring for instance the
AC current through the dot versus $\varepsilon_{d}$. \begin{figure}[h]
\includegraphics[width=1.\linewidth]{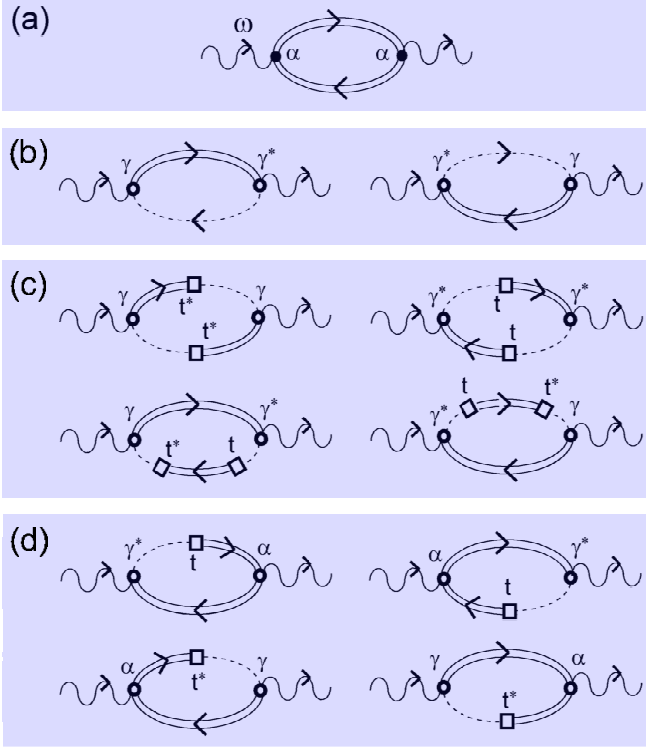}\label{diagrams}%
\caption{Scheme of the various contributions to $G_{\hat{h}_{int}\hat{h}%
_{int}}(t)$. The wavy lines correspond to photonic propagators, the simple
dashed lines to\ bare electronic propagators in the normal metal reservoir,
and the double full lines to electronic propagators in the quantum dot dressed
by the dot/reservoir tunneling processes.}%
\end{figure}

\subsection{Cavity frequency pull and damping pull}

Our purpose is to identify signatures of the terms in $\gamma$, which could
become visible in the cavity response, considering the recent progress on the
Mesoscopic QED technology\cite{Viennot2,Stehlik}. In the framework of the
linear response theory, the cavity electromagnetic response to $H_{d}$ is
determined by the Green's function\cite{DQDth} $\tilde{G}_{\hat{a},\hat
{a}^{\dag}}(\omega_{RF})$ with $\tilde{G}_{\hat{A},\hat{B}}(t)=-i\theta
(t)\left\langle [\hat{B}(t),\hat{A}(t=0)]\right\rangle $ and $\tilde{G}%
_{\hat{A},\hat{B}}(\omega)=%
%TCIMACRO{\tint \nolimits_{-\infty}^{+\infty}}%
%BeginExpansion
{\textstyle\int\nolimits_{-\infty}^{+\infty}}
%EndExpansion
dt\tilde{G}_{\hat{A},\hat{B}}(t)e^{i\omega t}$. One can use the exact
relation
\begin{equation}
\tilde{G}_{\hat{a},\hat{a}^{\dag}}(\omega)=\tilde{G}_{0}(\omega)+\hbar
^{-2}G_{0}(\omega)\tilde{G}_{\hat{h}_{int}\hat{h}_{int}}(\omega)G_{0}(\omega)
\end{equation}
with $G_{0}(\omega)=(\omega-\omega_{0}+i\Lambda_{0})^{-1}$ the free cavity
Green's function. In the simplest model where $\tau$ and $n_{0}$ are
independent of energy, the constant $\Lambda_{0}=\pi\left\vert \tau\right\vert
^{2}n_{0}/\hbar$ corresponds to pure cavity damping. In the following, we
consider the limit $\Lambda_{0}\ll$ $\omega_{0}$ which corresponds to standard
cavities. In the limit $\omega_{RF}-\omega_{0},\Lambda_{0}\ll\Gamma$, one can
use a random phase approximation which leads to
\begin{equation}
\tilde{G}_{\hat{a},\hat{a}^{\dag}}(\omega_{RF})\simeq\left(  \omega
_{RF}-\omega_{0}+i\Lambda_{0}-\hbar^{-2}G_{\hat{h}_{int}\hat{h}_{int}}%
(\omega_{RF})\right)  ^{-1} \label{rep}%
\end{equation}
with $G_{\hat{h}_{int}\hat{h}_{int}}(t)=\left.  \tilde{G}_{\hat{h}_{int}%
\hat{h}_{int}}(t)\right\vert _{\hat{h}_{int}=0}$ the electronic response
function calculated in the absence of the cavity. In the framework of a
non-interacting diagrammatic calculation, $G_{\hat{h}_{int}\hat{h}_{int}%
}(\omega)$ can be decomposed into the eleven contributions represented in Fig.
2. Our calculation is not perturbative in $t$ since we use an exact expression
for the dressed propagator of electrons on the dot, corresponding to the
double full lines in Fig.2. There are nine types of diagrams corresponding to
the possible ordered pairs formed with the interaction constants $\alpha$,
$\gamma$, and $\gamma^{\ast}$. The diagrams of Fig.2a correspond to pure
contributions from photon-induced orbital shifts in $\alpha^{2}$. The diagrams
of Figs. 2b and c correspond to contributions from photon-induced tunneling in
$\left\vert \gamma\right\vert ^{2}$, $\gamma^{2}$ or $\gamma^{\ast2}$. Note
that the contributions in $\left\vert \gamma\right\vert ^{2}$ have been
separated into the diagrams of Fig. 2b which diverge logarithmically for $D$
large and the bottom diagrams in Fig.2c which are regular. There also exist
interferences between photon-induced orbital shifts and photon-induced
tunneling, as shown by the diagrams of Fig. 2d which depend on $\alpha\gamma$
or $\alpha\gamma^{\ast}$.\ 

The constants $\Gamma$ and $D$ set the scale of variations of $G_{\hat
{h}_{int}\hat{h}_{int}}(\omega)$. Therefore, in the limit $\omega_{0}\ll
\Gamma,D$, one can replace $G_{\hat{h}_{int}\hat{h}_{int}}(\omega_{RF})$ by
$G_{\hat{h}_{int}\hat{h}_{int}}(0)$ in Eq.(\ref{rep}). This gives a cavity
frequency pull $\Delta\omega_{0}=G_{\hat{h}_{int}\hat{h}_{int}}(\omega
=0)/\hbar^{2}$ and a cavity damping pull $\Delta\Lambda_{0}=-(\omega
_{0}/i\hbar^{2})\left.  \partial G_{\hat{h}_{int}\hat{h}_{int}}(\omega
)/\partial\omega\right\vert _{\omega=0}$. In the limit $\Gamma\ll D$ and at
zero temperature, one obtains $\Delta\omega_{0}=\Delta\omega_{0}^{a}%
+\Delta\omega_{0}^{b}$ where%

\begin{align}
\Delta\omega_{0}^{a}  &  =\frac{2}{\pi\hbar(\Gamma^{2}+\varepsilon_{d}^{2}%
)}\label{a1}\\
&  \left(  4\pi^{2}N_{0}^{2}\operatorname{Re}[t\gamma^{\ast}]^{2}\Gamma
+4\pi\alpha\operatorname{Re}[t\gamma^{\ast}]N_{0}\varepsilon_{d}-\alpha
^{2}\Gamma\right) \nonumber
\end{align}
stems from the diagrams of Fig.2,a, c and d, and%
\begin{align}
\Delta\omega_{0}^{b}  &  =\frac{2}{\hbar}\left\vert \gamma\right\vert
^{2}N_{0}\frac{D}{D^{2}+\varepsilon_{d}^{2}}\label{a2}\\
&  \left(  D\log[\frac{\Gamma^{2}+\varepsilon_{d}^{2}}{D^{2}}]+\left\vert
\varepsilon_{d}\right\vert (2\arctan[\frac{\Gamma}{\left\vert \varepsilon
_{d}\right\vert }]-\pi)\right) \nonumber
\end{align}
stems from the diagrams of Fig.2,b. Note that the cutoff $\nu(\varepsilon)$ on
the reservoir density of states is necessary to avoid a divergence of
$\Delta\omega_{0}^{b}$. The cavity damping pull is given by%
\begin{equation}
\Delta\Lambda_{0}=\frac{2\omega_{0}}{\pi(\Gamma^{2}+\varepsilon_{d}^{2})^{2}%
}\left(  \alpha\Gamma-2\pi\operatorname{Re}[t\gamma^{\ast}]N_{0}%
\varepsilon_{d}\right)  ^{2}~\text{.} \label{pos}%
\end{equation}
Interestingly, $\Delta\omega_{0}^{a}$ and $\Delta\Lambda_{0}$ vanish for
$\varepsilon_{d}\gg\Gamma$, whereas $\Delta\omega_{0}^{b}$ vanishes only for
$\varepsilon_{d}\gg D$. Above, we have implicitly included the spin degree of
freedom in the model. In the spinless case, $\Delta\omega_{0}$ and
$\Delta\Lambda_{0}$ should be divided by a factor 2.

From Eqs.(\ref{a1}-\ref{pos}), at low temperatures, when the cavity has a
frequency smaller than the dot/reservoir tunnel rate $\Gamma$ and the
reservoir bandwidth $D$, and when $\gamma=0$, the cavity frequency pull and
the damping pull fulfill the relation $\Theta=\pi/2$ with%
\begin{equation}
\Theta=\frac{\alpha^{2}\Delta\Lambda_{0}}{\omega_{0}(\hbar\Delta\omega
_{0})^{2}}~\text{.}%
\end{equation}
This property is a manifestation of the Korringa-Shiba
relation\cite{KorringaShiba}, whose universality is still actively discussed
for quantum RC circuits \cite{RCbuttiker,RC1,RC2,RC3,RC4,RC5,RC6}. Remarkably,
we find that this universality is broken for $\gamma$ finite. Hence, one can
identify the existence of the $\gamma$ coupling by comparing the $\Delta
\omega_{0}(\varepsilon_{d})$ and $\sqrt{\Delta\Lambda_{0}}(\varepsilon_{d})$
curves (Fig. 3) or studying $\Theta$ (Fig. 4). Here, we present results for
$\gamma/\alpha\ll1$ because from Eqs. (\ref{jj1}-\ref{jj2}), this is the most
probable regime of parameters. When $\gamma=0$, $\Delta\omega_{0}$ and
$\sqrt{\Delta\Lambda_{0}}$ show identical variations with $\varepsilon_{d}$,
and are both even functions of $\varepsilon_{d}$, while $\Theta=\pi/2$. When
$\gamma\neq0$, $\sqrt{\Delta\Lambda_{0}}$ presents a resonance with
$\varepsilon_{d}$ which is wider than that of $\Delta\omega_{0}$ (Fig.3,
middle panel). A good resolution on the tails of the $\Delta\omega_{0}$ and
$\sqrt{\Delta\Lambda_{0}}$ resonances could also reveal that $\Delta\omega
_{0}$ and $\Delta\Lambda_{0}$ are not even with $\varepsilon_{d}$ in the
general case, except if $\arg[t\gamma^{\ast}]=\pi/2$ (Fig.3, bottom panel). In
principle, even a relatively small $\gamma$ can be detected. Indeed, for
$\gamma/\alpha\sim10^{-4}$, $\Theta$ already significantly deviates from
$\pi/2$ at $\varepsilon_{d}\sim0$ where the signals $\Delta\omega_{0}$ and
$\sqrt{\Delta\Lambda_{0}}$ are maximal (Fig.4, pink curves). This is because
the effect of the photon-induced tunneling term is boosted by the large number
of states involved in the reservoirs. To conclude, in order to reveal the
photon-induced tunneling terms, one promising possibility is to compare the
cavity frequency pull and damping pull caused by a quantum dot with a single
normal metal reservoir. Even a relatively weak photon-induced tunneling can
affect these quantities because the large number of reservoir states
reinforces the tunnel effect.\begin{figure}[h]
\includegraphics[width=0.7\linewidth]{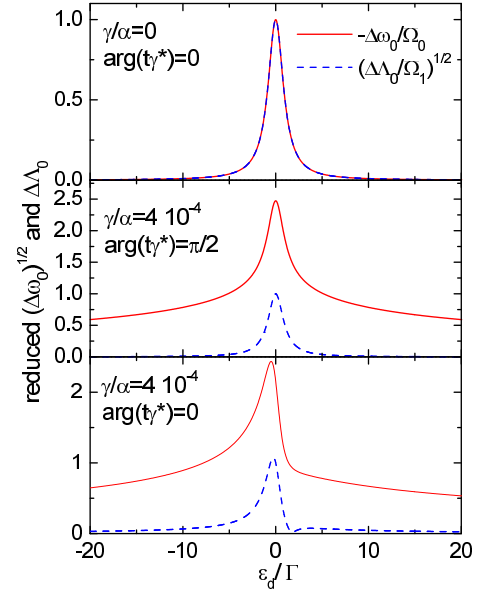}\caption{Reduced $\Delta
\omega_{0}$ and $\sqrt{\Delta\Lambda_{0}}$ versus $\varepsilon_{d}$ for
different values of $\gamma$. We have used $t/\Gamma=0.001$, $D=20\Gamma$, and
pulsation scales $\Omega_{0}=2\alpha^{2}/\pi\Gamma\hbar$ and $\Omega_{1}%
=\hbar\omega_{0}\Omega_{0}/\Gamma$.}%
\end{figure}\begin{figure}[h]
\includegraphics[width=0.7\linewidth]{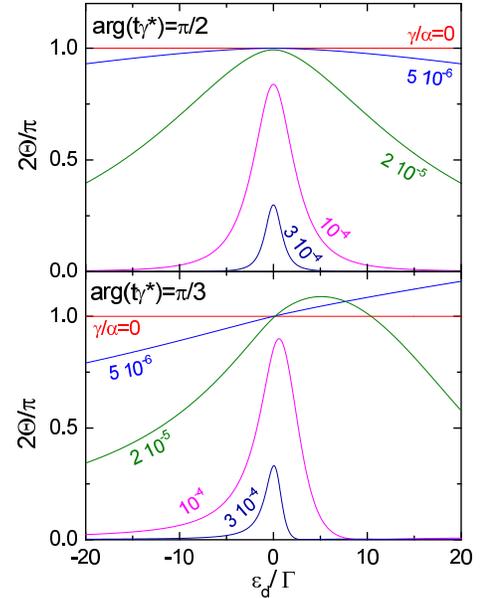}\caption{Ratio $\Theta$ versus
$\varepsilon_{d}$ for different values of $\gamma$. We have used the same
values of $t$, $D$ and $\Omega_{0}$ as in Fig.3}%
\end{figure}

\subsection{Experimental state of the art and discussion}

In principle, circuit-QED should enable an accurate determination of $\Theta$
thanks to the use of resonant techniques. The fabrication of a quantum dot
with a single normal metal reservoir inside a coplanar cavity is fully
accessible with present techniques. However, so far, the closest case studied
experimentally is a quantum dot with two normal metal reservoirs, which leads
to more complicated physics due to an asymmetric coupling of the two
reservoirs chemical potentials to the cavity\cite{Delbecq1,Delbecq2}.
Different types of quantum RC circuits based on 2-DEGs but also carbon
nanotubes or semiconducting nanowires deserve an investigation. The ratio
$\gamma/\alpha$ should be strongly dependent on the type of nanoconductor and
the contacts configuration used.

The physics discussed in this section is related to the problem of a quantum
RC circuit which is not coupled to a cavity but directly excited by a
classical microwave field with amplitude $a_{RF}$ and frequency $\omega
_{RF}/2\pi$, which corresponds to a term in $\hat{h}_{int}a_{RF}\cos
(\omega_{RF}t)$. At low frequencies and $T=0$, it has been
predicted\cite{RCbuttiker,RC1,RC2,RC3,RC4,RC5,RC6} that a spinless quantum RC
circuit excited classically should have the same admittance as an RC circuit,
with a universal resistance $R_{rel}=h/2e^{2}$. However, this result was
obtained in a purely capacitive coupling scheme i.e. $\gamma=0$. Assuming that
the admittance of the circuit is determined from the current $\hat{I}%
=ed\hat{n}_{d}/dt$ with $\hat{n}_{d}=\hat{c}_{d}^{\dag}\hat{c}_{d}$, this
value of $R_{rel}$ can be recovered by calculating the charge susceptibility
$G_{\hat{n}_{d},\hat{n}_{d}}(\omega_{RF})$, i.e. the diagram of Fig. 2.a,
which leads to $R_{rel}=\Theta\hbar/e^{2}=h/2e^{2}$ because $\gamma=0$.
However, in the general case, one should calculate the current response of the
quantum RC circuit from $G_{\hat{n}_{d},\hat{h}_{int}}(\omega_{RF})$ because
the oscillating drive can also modify the tunnel barrier transparency. Only
the diagram of Fig. 2.a and half of the diagrams of Fig. 2d should contribute
to $R_{rel}$, whereas the diagrams of Fig.2.b leading to the logarithmic term
of Eq.(\ref{a2}) should not contribute. Therefore, $R_{rel}$ should show a
behavior qualitatively different from $\Theta$. We will not discuss the
detailed behavior of $R_{rel}$ here. We will simply point out that for
$\gamma\neq0$, $R_{rel}$ is not universal, and furthermore, $R_{rel}$ and
$\Theta$ are not trivially related. Interestingly, the universal value
$R_{rel}=h/2e^{2}$ expected for a spinless system with $\gamma=0$ has been
checked experimentally in the limit $T\rightarrow0$ for a 2-DEG based
circuit\cite{Gabelli}. However, this measurement is not incompatible with a
finite $\gamma$, due to its uncertainty of the order of $\pm20\%$%
\cite{theseGabelon}.

Note that since $R_{rel}$ and $\Theta$ are qualitatively different signals, it
can be interesting to measure both in the same experiment in order to obtain
more information on the system. More generally, the study of tunneling effects
can benefit from a simultaneous measurement of the cavity response and the
current through the nanocircuit. Such a joint measurement has already been
realized in the case of a DQD. The cavity dispersive shift is directly
sensitive to the population imbalance between the DQD bounding and
antibounding states whereas the current though the DQD corresponds to a more
complex combination of state populations and tunnel rates. The simultaneous
measurement of the two signals gives stronger constraints to determine the
system parameters\cite{Viennot}. Another possibility to gain more information
on Mesoscopic QED systems might be to measure cross-correlations between the
electronic current and the cavity output field.

Importantly, the example of the quantum RC circuit shows that dissipation of
the cavity photons has to be considered carefully, since an open quantum dot
circuit with a single contact can already induce photon dissipation. On a more
general level, photonic dissipation in Mesoscopic QED is a very rich problem.
Another interesting possibility would be to prepare non-trivial quantum
photonic states by combining the effects of photonic and electronic
dissipation with the non-linearity of quantum dot circuits, by analogy with
the methods developped for two level systems\cite{Sarlette,Holland}. This
so-called "reservoir engineering" could benefit from the specificities of
nanocircuits like for instance the use of reservoirs with specific electronic
orders or finite bias electronic transport .

\section{Conclusion}

Mesoscopic QED has common ingredients with atomic Cavity QED, like the
relevance of an orbital degree of freedom, which does not exist for
superconducting qubits. It has also common ingredients with standard Circuit
QED, like the tunneling physics and the strong inhomogeneities of the photonic
modes. It is therefore necessary to develop a specific description which
combines all these ingredients. In the case where photon induced magnetic
effects can be disregarded, one can express the coupling between tunneling
quasiparticles and photons in terms of a scalar photonic pseudo-potential. In
the framework of a tunneling model, this leads to photon-induced orbital
energy shifts, which coexist with photon-induced tunneling terms and
photon-induced local orbital transitions.\textbf{ }To illustrate the richness
of our approach, we have discussed the example of a cavity coupled to a
quantum RC circuit, i.e. a single quantum dot coupled to a single normal metal
reservoir. The photon-induced tunneling terms between the dot and the
reservoir induce a non-universal relation between the cavity frequency pull
and the cavity damping pull, contrarily to what is expected with purely
capacitive coupling schemes at low
temperatures\cite{RCbuttiker,RC1,RC2,RC3,RC4,RC5,RC6}. This case represents
only one example of use for our formalism. We have given explicit Hamiltonians
for the cases of a multi-quantum dot circuit in a cavity, and a
superconducting nanostructure enclosing Majorana bound states coupled to a
cavity. One could also consider cavities coupled to quantum point contacts,
molecular circuits, and metallic tunnel junctions, for instance.

\textit{We acknowledge useful discussions with M. Baillergeau, L. E. Bruhat,
C. Cohen-Tannoudji, M. C. Dartiailh, M. R. Delbecq, M. P. Desjardins, R. B.
Saptsov, M. Trif, and J. J. Viennot. This work was financed by the ERC
Starting grant CirQys, the EU FP7 project SE2ND[271554] and the
ANR-NanoQuartet [ANR12BS1000701] (France).}

\section{Appendix A: Effective model for Mesoscopic QED devices}

\subsection{Definition of the model}

\textit{ }\begin{figure}[h]
\includegraphics[width=1.\linewidth]{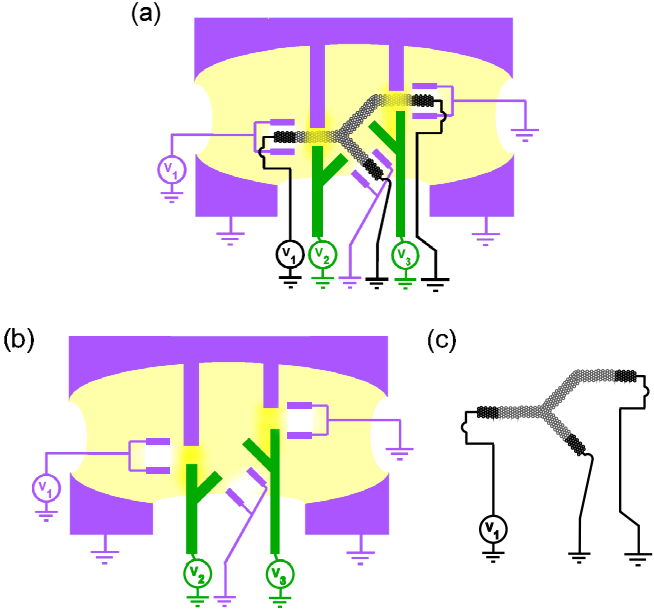}\caption{(a) Effective
model for the circuit of Fig.1, where the nanocircuit reservoirs are
decomposed into effective orbital reservoirs (dark grey elements) and
effective plasmonic reservoirs (nearby purple elements) (b) Ensemble
$\mathcal{C}$ of the perfect conductors considered for the generation of the
photonic modes. The photonic field in yellow is inhomogeneous, as represented
by darker yellow areas near the cavity protusions, and by the white
(screening) areas between the small purple conductors (c) Effective orbital
nanocircuit $\mathcal{O}$.}%
\label{CircuitEq}%
\end{figure}In the main text, Eqs (\ref{1})-(\ref{6}) have been introduced on
the basis of a physical discussion and gauge invariance considerations. It is
also instructive to use an effective model which separates spatially the
tunneling and plasmonic parts of the nanocircuit. Note that this effective
model does not aim at calculating quantitatively the nanocircuit/cavity
couplings. It rather aims at providing a full mathematical justification for
the form of Eqs. (\ref{1})-(\ref{6}) and a deeper insight on the underlying
physics. For simplicity, we assume that the nanoconductors host only
individual electronic orbital modes since their electronic density is
generally very low. We assume that the plasmonic modes in the fermionic
reservoirs (blue elements in Fig. 1) are ultra fast due to the absence of
dynamical Coulomb blockade\cite{DCB}. In this case, one can decompose each
fermionic reservoir into a purely orbital conductor (dark grey elements in
Fig.5.a) which contains no plasmons and is tunnel coupled to the nanocircuit,
and perfect conductors (small purple elements in Fig.5.a) which can host
plasmonic modes, but are only capacitively coupled to the nanoconductors.
These perfect conductors, or "effective plasmonic reservoirs", can be placed
such that the partial or full decrease of the photonic fields around the
nanocircuit orbital states is reproduced at least qualitatively. We denote
with $\mathcal{C}$ the ensemble of the perfect conductors, which includes the
cavity conductors, gate electrodes, and the effective plasmonic reservoirs
(see Fig. 5.b). Some of these conductors are voltage biased and some others
are left floating, like for instance the central conductor in a coplanar
cavity. The rest of the circuit represents an effective orbital nanocircuit
$\mathcal{O}$ where tunneling physics prevails (see Fig. 5.c). Note that in
principle, to grant current conservation, the "orbital" reservoirs have to be
connected to voltage sources through wirings which host plasmonic modes, but
one can assume that they are too far from the nanoconductors to have any
significant influence on the nanocircuit/cavity coupling.

For later use, we attribute to each conductor of the ensemble $\mathcal{C}$ an
index $i$ and we note $\mathcal{S}_{i}$ and $\mathcal{V}_{i}$ the
corresponding surface and volume. In the following treatment we make a
distinction between the conductors $i\in\mathbb{F}$ which are left floating
with a constant total charge $Q_{i}$, and the conductors $i\in\mathbb{B}$
which are voltage-biased with a constant voltage $V_{i}$ or grounded
($V_{i}=0$). The generators imposing the voltages $V_{i}$ can be omitted from
the description and included in electromagnetic boundary conditions since they
are far from the cavity and nanocircuits. The volume $\mathcal{V}%
_{\mathcal{O}}$ complementary to $%
%TCIMACRO{\tsum _{i}}%
%BeginExpansion
{\textstyle\sum_{i}}
%EndExpansion
\mathcal{V}_{i}$ hosts a charge distribution $\rho(\overrightarrow{r},t)=%
%TCIMACRO{\tsum \nolimits_{\alpha\in\mathcal{O}}}%
%BeginExpansion
{\textstyle\sum\nolimits_{\alpha\in\mathcal{O}}}
%EndExpansion
e_{\alpha}\delta(\overrightarrow{r}-\overrightarrow{q}_{\alpha})$ of particles
$\alpha$ with charges $e_{\alpha}$ at positions $\overrightarrow{q}_{\alpha}$,
which belong to the nanocircuit $\mathcal{O}$. We also define the
corresponding current distribution $\overrightarrow{j}(\overrightarrow{r},t)=%
%TCIMACRO{\tsum \nolimits_{\alpha\in\mathcal{O}}}%
%BeginExpansion
{\textstyle\sum\nolimits_{\alpha\in\mathcal{O}}}
%EndExpansion
e_{\alpha}\overrightarrow{\dot{q}}_{\alpha}\delta(\overrightarrow
{r}-\overrightarrow{q}_{\alpha})$.

\subsection{Decomposition of fields in harmonic, longitudinal, and transverse
components.}

In Cavity QED, the treatment of the boundary conditions provided by the cavity
conductors is usually omitted, because the atoms are very far from the cavity
mirrors. Therefore, a procedure based on a spatial Fourier transformation can
be used to separate the transverse electric field $\overrightarrow{E}_{\perp}%
$, which has no divergence, from the longitudinal electric field
$\overrightarrow{E}_{\parallel}$, which has no
rotational\cite{Cohen-Tannoudji}. It is found that the value\ of
$\overrightarrow{E}_{\parallel}$ is instantaneously imposed by the charge
distribution $\rho(\overrightarrow{r},t)$ associated to the atomic system. In
contrast, the transverse fields $\overrightarrow{E}_{\perp}$ and
$\overrightarrow{B}=\overrightarrow{B}_{\perp}$ correspond to propagating
modes. In Mesoscopic QED, it is necessary generalize this procedure to account
for the presence of the perfect conductors of the ensemble $\mathcal{C}$,
including voltage-biased electrostatic gates. As already mentioned in the
previous section, we use a charge distribution $\rho(\overrightarrow{r},t)$
which includes the charges from the $\mathcal{O}$ subsystem only. We will take
into account the screening charges on the surface of the conductors of
$\mathcal{C}$ through electromagnetic boundary conditions.\ 

Inside $V_{\mathcal{O}}$, $\rho(\overrightarrow{r},t)$, $\overrightarrow
{j}(\overrightarrow{r},t)$, and the total electric and magnetic fields
$\overrightarrow{E}(\overrightarrow{r},t)$ and $\overrightarrow{B}%
(\overrightarrow{r},t)$ are related by Maxwell's equations. To understand how
these relations are affected by the screening charges lying on $\mathcal{C}$,
we introduce the Hodge decomposition of a generic field $\vec{f}$ as $\vec
{f}=\vec{f}_{\parallel}+\vec{f}_{\perp}+\vec{f}_{harm}$, where $\vec
{f}_{\parallel}$ has a finite divergence but no rotational, $\vec{f}_{\perp}$
has a finite rotational but no divergence, and $\vec{f}_{harm}$ has
none\cite{Hodge,Hodge2}. A similar decomposition was invoked very recently in
the context of Cavity QED \cite{Vukics}. However, in this Ref., the effects of
the DC voltage biases $V_{i}$ or the floating charges $Q_{i}$ are not
considered (i.e. $V_{i}=0$ is used for any $i$). Besides, the volume
$V_{\mathcal{O}}$ outside the cavity conductors is assumed to be bounded. This
hypothesis is not possible in our case due to openings in the cavity planes,
which are necessary to connect the nanocircuits to their outside voltage bias
circuitry. Hence, we need to perform the Hodge decomposition in the fully
general case. Below, we explain how to perform this task.

It is convenient to introduce a potential $\Phi_{harm}(\overrightarrow{r})$
and a Green's function $G(\overrightarrow{r},\overrightarrow{r}^{\prime})$
which fulfill the static Laplace equations:%
\begin{equation}
\Delta_{\overrightarrow{r}}\Phi_{harm}(\overrightarrow{r})=0 \label{bbc0}%
\end{equation}
and
\begin{equation}
\Delta_{\overrightarrow{r}}G(\overrightarrow{r},\overrightarrow{r}^{\prime
})=-\delta(\overrightarrow{r}-\overrightarrow{r}^{\prime})/\varepsilon_{0}
\label{bc0}%
\end{equation}
for $(\overrightarrow{r},\overrightarrow{r}^{\prime})\in\mathcal{V}%
_{\mathcal{O}}^{2}$, with boundary conditions%
\begin{equation}%
%TCIMACRO{\tint \nolimits_{\mathcal{S}_{i}}}%
%BeginExpansion
{\textstyle\int\nolimits_{\mathcal{S}_{i}}}
%EndExpansion
d^{2}r\overrightarrow{\nabla}_{\overrightarrow{r}}\Phi_{harm}(\overrightarrow
{r}).\vec{n}=-Q_{i}\text{ for }i\in\mathbb{F}~\text{,}%
\end{equation}%
\begin{equation}
\Phi_{harm}(\overrightarrow{r})=V_{i}\text{ for }\overrightarrow{r}%
\in\mathcal{S}_{i}\text{ and }i\in\mathbb{B}%
\end{equation}
and
\begin{equation}%
%TCIMACRO{\tint \nolimits_{\mathcal{S}_{i}}}%
%BeginExpansion
{\textstyle\int\nolimits_{\mathcal{S}_{i}}}
%EndExpansion
d^{2}r\overrightarrow{\nabla}_{\overrightarrow{r}}G(\overrightarrow
{r},\overrightarrow{r}^{\prime}).\vec{n}=0\text{ for }i\in\mathbb{F}~\text{,}
\label{bc1}%
\end{equation}%
\begin{equation}
G(\overrightarrow{r},\overrightarrow{r}^{\prime})=0\text{ for }\overrightarrow
{r}\in\mathcal{S}_{i}\text{ and }i\in\mathbb{B}~\text{.} \label{bc2}%
\end{equation}
Above, $\vec{n}$ is the outward pointing unit vector perpendicular to
$\mathcal{S}_{i}$. One can check that each of these two sets of Eqs. has a
unique solution provided $\mathbb{B}$ is not empty\cite{Jackson}.\textbf{
}From Eqs. (\ref{bbc0}-\ref{bc2}) and the Maxwell Eqs., one can check that:%
\begin{equation}
\vec{E}_{harm}(\overrightarrow{r})=-\overrightarrow{\nabla}_{\overrightarrow
{r}}\Phi_{harm}(\overrightarrow{r})~\text{,} \label{d0}%
\end{equation}%
\begin{equation}
\vec{E}_{\parallel}(\overrightarrow{r},t)=-%
%TCIMACRO{\tint }%
%BeginExpansion
{\textstyle\int}
%EndExpansion
d^{3}r^{\prime}\overrightarrow{\nabla}_{\overrightarrow{r}}G(\overrightarrow
{r},\overrightarrow{r}^{\prime})\rho(\overrightarrow{r}^{\prime},t) \label{d1}%
\end{equation}
and%
\begin{equation}
\vec{E}_{\perp}(\overrightarrow{r},t)=\vec{E}(\overrightarrow{r},t)-\vec
{E}_{\parallel}(\overrightarrow{r},t)-\vec{E}_{harm}(\overrightarrow{r})
\label{d2}%
\end{equation}
while $\vec{B}_{\perp}(\overrightarrow{r},t)=\vec{B}(\overrightarrow{r},t)$ is
still valid. We also find $\overrightarrow{j}=\overrightarrow{j}_{\parallel
}+\overrightarrow{j}_{\perp}$ with%
\begin{equation}
\overrightarrow{j}_{\parallel}(\overrightarrow{r},t)=-(\partial\overrightarrow
{E}_{\parallel}(\overrightarrow{r},t)/\partial t)/c^{2}\mu_{0} \label{d3}%
\end{equation}
and%
\begin{equation}
\overrightarrow{j}_{\perp}(\overrightarrow{r},t)=\overrightarrow
{j}(\overrightarrow{r},t)-\overrightarrow{j}_{\parallel}(\overrightarrow
{r},t)~\text{.} \label{d4}%
\end{equation}
The harmonic component $\vec{E}_{harm}$ represents a static contribution to
the electric field, set by the charges $Q_{i}$ or potentials $V_{i}$ imposed
on the conductors $i$. Thanks to the use of $G(\overrightarrow{r}%
,\overrightarrow{r}^{\prime})$, the definitions (\ref{d1}-\ref{d4}) take into
account the effect of the screening charges lying on $\mathcal{C}$. Using the
Maxwell equations and the expressions (\ref{d0}-\ref{d4}), one can check that
the longitudinal components $\vec{E}_{\parallel}$ and $\overrightarrow
{j}_{\parallel}$ are set instantaneously by $\rho(\overrightarrow{r},t)$
since
\begin{equation}
\overrightarrow{\nabla}_{\overrightarrow{r}}.\vec{E}_{\parallel}%
(\overrightarrow{r},t)=\rho(\overrightarrow{r},t)/\varepsilon_{0} \label{b1}%
\end{equation}
for $\overrightarrow{r}\in\mathcal{V}_{\mathcal{O}}$. Furthermore,
$\overrightarrow{B}_{\perp}$, $\overrightarrow{E}_{\perp}$ and
$\overrightarrow{j}_{\perp}$ follow an independent system of propagation
equations, i.e.
\begin{equation}
\overrightarrow{\nabla}_{\overrightarrow{r}}\wedge\overrightarrow{E}_{\perp
}(\overrightarrow{r},t)=-\partial\overrightarrow{B}_{\perp}(\overrightarrow
{r},t)/\partial t~\text{,} \label{b3}%
\end{equation}%
\begin{equation}
\overrightarrow{\nabla}_{\overrightarrow{r}}\wedge\overrightarrow{B}_{\perp
}(\overrightarrow{r},t)=\mu_{0}\overrightarrow{j}_{\perp}(\overrightarrow
{r},t)+\frac{1}{c^{2}}\frac{\partial}{\partial t}\overrightarrow{E}_{\perp
}(\overrightarrow{r},t) \label{b4}%
\end{equation}
for $\overrightarrow{r}\in\mathcal{V}_{\mathcal{O}}$. The above equations have
to be supplemented with the boundary conditions%
\begin{equation}%
%TCIMACRO{\tint \nolimits_{\mathcal{S}_{i}}}%
%BeginExpansion
{\textstyle\int\nolimits_{\mathcal{S}_{i}}}
%EndExpansion
d^{2}r\vec{E}_{\parallel(\perp)}(\overrightarrow{r},t).\vec{n}=0\text{ for
}i\in\mathbb{F}~\text{,} \label{bb1}%
\end{equation}%
\begin{equation}%
%TCIMACRO{\tint \nolimits_{\ell_{ij}}}%
%BeginExpansion
{\textstyle\int\nolimits_{\ell_{ij}}}
%EndExpansion
\overrightarrow{dr}.\vec{E}_{\parallel(\perp)}(\overrightarrow{r},t)=0\text{
for }\overrightarrow{r}\in\mathcal{S}_{i}\text{ with }(i,j)\in\mathbb{B}^{2}
\label{bb2}%
\end{equation}
with $\ell_{ij}$ any trajectory connecting conductors $i$ and $j$.

\subsection{Mesoscopic QED classical Lagrangian\textit{.}\label{Lagrangian}}

At this stage, it is convenient to introduce scalar and vector potentials $U$
and $\vec{A}$ in the Coulomb gauge i.e. $\overrightarrow{\nabla}%
_{\overrightarrow{r}}.\vec{A}(\overrightarrow{r},t)=0$. We also define the
decomposition $U(\overrightarrow{r},t)=U_{\parallel}(\overrightarrow
{r},t)+\Phi_{harm}(\overrightarrow{r})$, such that%
\begin{equation}
\vec{E}_{\parallel}(\overrightarrow{r},t)=-\overrightarrow{\nabla
}_{\overrightarrow{r}}U_{\parallel}(\overrightarrow{r},t)~\text{,} \label{l2}%
\end{equation}%
\begin{equation}
\overrightarrow{E}_{\perp}(\overrightarrow{r},t)=-\partial\vec{A}%
(\overrightarrow{r},t)/\partial t~\text{,} \label{l3}%
\end{equation}
and%
\begin{equation}
\vec{B}=\overrightarrow{\nabla}_{\overrightarrow{r}}\wedge\vec{A}%
(\overrightarrow{r},t)~\text{.} \label{l4}%
\end{equation}
A comparison between Eqs. (\ref{bc0}-\ref{bc2}), and (\ref{b1},\ref{bb1}%
,\ref{bb2}) leads to the identification
\begin{equation}
U_{\parallel}(\overrightarrow{r},t)=%
%TCIMACRO{\tint _{\mathcal{V}_{\mathcal{O}}}}%
%BeginExpansion
{\textstyle\int_{\mathcal{V}_{\mathcal{O}}}}
%EndExpansion
d^{3}r^{\prime}G(\overrightarrow{r},\overrightarrow{r}^{\prime})\rho
(\overrightarrow{r}^{\prime},t) \label{l1}%
\end{equation}
up to a global constant which one can disregard. Hence, $\vec{E}_{\parallel}$
but also $U_{\parallel}$ are instantaneously set by the charge distribution
$\rho(\overrightarrow{r},t)$. In contrast, the transverse fields
$\overrightarrow{E}_{\perp}$ and $\vec{B}$ are determined by $\vec{A}$ which
follows the propagation equation
\begin{equation}
\square_{\overrightarrow{r}}(\vec{A}(\overrightarrow{r},t))=-\mu
_{0}\overrightarrow{j}_{\perp}(\overrightarrow{r},t) \label{field}%
\end{equation}
given by Eq.(\ref{b4}), with $\square_{\overrightarrow{r}}=\Delta
_{\overrightarrow{r}}-(\partial^{2}/c^{2}\partial t^{2})$, and boundary
conditions similar to Eqs.(\ref{bb1})-(\ref{bb2}). In this framework, a proper
classical Lagrangian is
\begin{align}
L  &  =\frac{1}{2}%
%TCIMACRO{\tsum \nolimits_{\alpha}}%
%BeginExpansion
{\textstyle\sum\nolimits_{\alpha}}
%EndExpansion
m_{\alpha}\dot{q}_{\alpha}^{2}-\frac{1}{2}%
%TCIMACRO{\tsum \nolimits_{\alpha}}%
%BeginExpansion
{\textstyle\sum\nolimits_{\alpha}}
%EndExpansion
e_{\alpha}U_{\parallel}(\overrightarrow{q}_{\alpha},t)\label{LL}\\
&  -%
%TCIMACRO{\tsum \nolimits_{\alpha}}%
%BeginExpansion
{\textstyle\sum\nolimits_{\alpha}}
%EndExpansion
e_{\alpha}\Phi_{harm}(\overrightarrow{q}_{\alpha})+%
%TCIMACRO{\tsum \nolimits_{\alpha}}%
%BeginExpansion
{\textstyle\sum\nolimits_{\alpha}}
%EndExpansion
e_{\alpha}\overrightarrow{\dot{q}}_{\alpha}.\overrightarrow{A}(\overrightarrow
{q}_{\alpha},t)\nonumber\\
&  +\frac{\varepsilon_{0}}{2}%
%TCIMACRO{\tint _{\mathcal{V}_{\mathcal{O}}}}%
%BeginExpansion
{\textstyle\int_{\mathcal{V}_{\mathcal{O}}}}
%EndExpansion
d^{3}r\left(  \left\vert \frac{d}{dt}\overrightarrow{A}(\overrightarrow
{r},t)\right\vert ^{2}-c^{2}\left\vert \overrightarrow{\nabla}%
_{\overrightarrow{r}}\wedge\overrightarrow{A}(\overrightarrow{r},t)\right\vert
^{2}\right) \nonumber
\end{align}
with $\alpha\in\mathcal{O}$ in the sums. The Lagrangian equations deriving
from $L$ correspond to Eq.(\ref{field}) and to the standard Newton equations
for the motion of particles $\alpha$.

\subsection{Mesoscopic QED Hamiltonian}

From Eq. (\ref{LL}), one can check that $\overrightarrow{p}_{\alpha}%
=m_{\alpha}\overrightarrow{\dot{q}}_{\alpha}+e_{\alpha}\overrightarrow
{A}(\overrightarrow{q}_{\alpha},t)$ is the conjugate of $\overrightarrow
{q}_{\alpha}$ and $\vec{\Pi}_{\perp}(\overrightarrow{r},t)=-\varepsilon
_{0}E_{\perp}(\overrightarrow{r},t)$ is the conjugate of $\overrightarrow
{A}(\overrightarrow{r},t)$. Therefore one can express the Hamiltonian of the
system as $H=%
%TCIMACRO{\tsum \nolimits_{\alpha}}%
%BeginExpansion
{\textstyle\sum\nolimits_{\alpha}}
%EndExpansion
\overrightarrow{p}_{\alpha}.\overrightarrow{\dot{q}}_{\alpha}+%
%TCIMACRO{\tint _{\mathcal{V}_{\mathcal{O}}}}%
%BeginExpansion
{\textstyle\int_{\mathcal{V}_{\mathcal{O}}}}
%EndExpansion
d^{3}r(\vec{\Pi}_{\perp}(\overrightarrow{r},t).\overrightarrow{\dot{A}%
}(\overrightarrow{r},t))-L$. This Hamiltonian can be quantized
straightforwardly by replacing $\overrightarrow{p}_{\alpha}$, $\overrightarrow
{q}_{\alpha}$, $\overrightarrow{\Pi}$ and $\overrightarrow{A}$ by operators
$\overrightarrow{\hat{p}}_{\alpha}$, $\overrightarrow{\hat{q}}_{\alpha}$,
$\overrightarrow{\hat{\Pi}}$ and $\overrightarrow{\hat{A}}$. This gives%
\begin{align}
\hat{H}  &  =\frac{1}{2m_{\alpha}}%
%TCIMACRO{\tsum \nolimits_{\alpha}}%
%BeginExpansion
{\textstyle\sum\nolimits_{\alpha}}
%EndExpansion
(\overrightarrow{\hat{p}}_{\alpha}-e_{\alpha}\overrightarrow{\hat{A}%
}(\overrightarrow{\hat{q}}_{\alpha}))^{2}\\
&  +\frac{1}{2}%
%TCIMACRO{\tsum \nolimits_{\alpha}}%
%BeginExpansion
{\textstyle\sum\nolimits_{\alpha}}
%EndExpansion
e_{\alpha}U_{\parallel}(\overrightarrow{\hat{q}}_{\alpha})+%
%TCIMACRO{\tsum \nolimits_{\alpha}}%
%BeginExpansion
{\textstyle\sum\nolimits_{\alpha}}
%EndExpansion
e_{\alpha}\Phi_{harm}(\overrightarrow{\hat{q}}_{\alpha})\nonumber\\
&  +\frac{1}{2}%
%TCIMACRO{\tint _{\mathcal{V}_{\mathcal{O}}}}%
%BeginExpansion
{\textstyle\int_{\mathcal{V}_{\mathcal{O}}}}
%EndExpansion
d^{3}r\left(  \frac{1}{\varepsilon_{0}}\left\vert \overrightarrow{\hat{\Pi}%
}\right\vert ^{2}+\frac{1}{\mu_{0}}\left\vert \overrightarrow{\nabla
}_{\overrightarrow{r}}\wedge\overrightarrow{\hat{A}}(\overrightarrow
{r})\right\vert ^{2}\right)  ~\text{.}\nonumber
\end{align}
In principle, the ensemble of the charges $\alpha\in\mathcal{O}$ includes
tunneling electrons ($\alpha\in\mathcal{T}$) but also ions and valence
electrons from an underlying crystalline structure. In the main text, the two
latter are assumed to be decoupled from the cavity and are thus treated in a
mean field approach. Note that the harmonic potential term in the above
Hamiltonian is omitted by Ref. \cite{Vukics}. In our case, this term is
crucial to account for the effect of DC electrostatic gates.

Assuming that the coupling between $\mathcal{C}$ and $\mathcal{O}$ is
perturbative, the fields $\overrightarrow{\hat{A}}$ and $\overrightarrow
{\hat{\Pi}}$ can be expressed in terms of photonic modes calculated in the
absence of $O$, using for instance Eq.(\ref{field}) with $\overrightarrow
{j}_{\perp}=0$ \ We treat explicitly only one of these modes in Eq.(\ref{6}).
Screening charges on the nanocircuits reservoirs can strongly modify
$\overrightarrow{\hat{A}}$. This is why, in the general case, it is not
adequate to make a perturbative treatment between the photonic field generated
by the empty cavity and the whole nanocircuit. Instead, the present work
considers a perturbative treatment between the photonic field generated by the
plasmonic modes in the whole Mesoscopic QED device, and the effective orbital
nanocircuit $\mathcal{O}$. Similarly, in superconducting Circuit QED, the
cavity modes can be strongly renormalized by the presence of a superconducting
quantum bit, and perturbation schemes must therefore be defined
carefully\cite{Nigg}.

In this Appendix, the renormalization of the bare cavity modes by the
nanocircuit plasmons is described formally by supplementary boundary
conditions imposed by effective plasmonic reservoirs. This enables a simple
mathematical justification for the form of Eqs (\ref{1})-(\ref{6}). In order
to get realistic estimates of the fields $\overrightarrow{\hat{A}}$ and
$\overrightarrow{\hat{\Pi}}$, one can make a numerical microwave simulation
using the real device geometry, with the nanoconductors omitted (see main text).

\section{Appendix B: Advantages of the photonic pseudo-potential scheme}

The advantages of a formalism without the $\hat{A}^{2}$ terms have already
been discussed thoroughly for Cavity QED\cite{Cohen-Tannoudji}. However, since
the physical ingredients and measured quantities are different in Cavity and
Mesoscopic QED, it is useful to discuss more specifically the advantages of
the photonic pseudo-potential scheme of section \ref{PPP}.

The first quantity which is generally measured in Mesoscopic QED experiments
is the cavity frequency pull for a given cavity mode, which is modified by the
nanocircuits coupled to the cavity. Since the minimal coupling scheme and the
photonic pseudo-potential scheme are related by a unitary transformation, they
must of course predict the same cavity pull. It is nevertheless instructive to
check this equivalence for a simple example. Here, we use a simpler approach
than in section IV, because we do not need to particularize the effect of the
nanocircuit reservoirs and we will not discuss the cavity damping pull. We
consider a cavity coupled to a nanocircuit with unperturbed many electron
states $\left\vert \varphi_{j}\right\rangle $ which satisfy the general
equation $\hat{H}_{0}\left\vert \varphi_{j}\right\rangle =E_{j}\left\vert
\varphi_{j}\right\rangle $ with%
\begin{equation}
\hat{H}_{0}=%
%TCIMACRO{\dsum \limits_{\alpha}}%
%BeginExpansion
{\displaystyle\sum\limits_{\alpha}}
%EndExpansion
\left(  \frac{\hat{p}_{\alpha}^{2}}{2m_{\alpha}}-e\Phi_{tot}(\overrightarrow
{\hat{q}}_{\alpha})\right)  +\frac{e^{2}}{2}%
%TCIMACRO{\dsum \limits_{\alpha,\alpha^{\prime}}}%
%BeginExpansion
{\displaystyle\sum\limits_{\alpha,\alpha^{\prime}}}
%EndExpansion
G(\overrightarrow{\hat{q}}_{\alpha},\overrightarrow{\hat{q}}_{\alpha^{\prime}%
})
\end{equation}
and $\Phi_{tot}(\overrightarrow{\hat{q}}_{\alpha})=\Phi_{harm}(\overrightarrow
{\hat{q}}_{\alpha})+V_{conf}(\overrightarrow{\hat{q}}_{\alpha})$. We discuss
the limit where the nanocircuit and the cavity are not resonant i.e.
$E_{j}-E_{j^{\prime}}\neq\hbar\omega_{0}$ for any $j$ and $j^{\prime}$. At
second order in the cavity/nanocircuit coupling (i.e. $V_{\perp}$ or
$\mathcal{\vec{A}}$), an elementary perturbation theory gives the cavity pull
\begin{equation}
\Delta\hbar\omega_{0,j}=%
%TCIMACRO{\dsum \limits_{j^{\prime},\alpha,\alpha^{\prime}}}%
%BeginExpansion
{\displaystyle\sum\limits_{j^{\prime},\alpha,\alpha^{\prime}}}
%EndExpansion
\frac{2(\hbar\omega_{0})^{2}(E_{j}-E_{j^{\prime}})\mathcal{M}_{jj^{\prime}%
}^{\alpha\alpha^{\prime}}}{((E_{j}-E_{j^{\prime}})^{2}-(\hbar\omega_{0})^{2})}
\label{pull}%
\end{equation}
with%
\begin{equation}
\mathcal{M}_{jj^{\prime}}^{\alpha\alpha^{\prime}}=\left\langle \varphi
_{j}\right\vert V_{\perp}(\overrightarrow{\hat{q}_{\alpha}})\left\vert
\varphi_{j^{\prime}}\right\rangle \left\langle \varphi_{j^{\prime}}\right\vert
V_{\perp}(\overrightarrow{\hat{q}_{\alpha^{\prime}}})\left\vert \varphi
_{j}\right\rangle ~\text{.}%
\end{equation}
Note that $\Delta\hbar\omega_{0,j}$ depends on the state $\left\vert
\varphi_{j}\right\rangle $ occupied by the nanocircuit. This can be used, in
principle, for a noninvasive readout of the nanocircuit state in the
non-resonant regime\cite{Wallraff}. Equation (\ref{pull}) can be obtained by
using Hamiltonian (\ref{Htottild}) or equivalently Hamiltonian (\ref{1}). In
order to check the equivalence between the two approaches, it is necessary to
invoke the completeness relation $%
%TCIMACRO{\tsum \nolimits_{j}}%
%BeginExpansion
{\textstyle\sum\nolimits_{j}}
%EndExpansion
\left\vert \varphi_{j}\right\rangle \left\langle \varphi_{j}\right\vert =1$
for the nanocircuit states. Interestingly, one can check that the $\hat{A}%
^{2}$ terms in Eq.(\ref{1}) give a contribution
\begin{equation}
\Delta\hbar\omega_{0,j}^{2ph}=%
%TCIMACRO{\dsum \limits_{j^{\prime},\alpha,\alpha^{\prime}}}%
%BeginExpansion
{\displaystyle\sum\limits_{j^{\prime},\alpha,\alpha^{\prime}}}
%EndExpansion
2(E_{j}-E_{j^{\prime}})\mathcal{M}_{jj^{\prime}}^{\alpha\alpha^{\prime}}%
\end{equation}
to $\Delta\hbar\omega_{0,j}$, which is not negligible for $\left\vert
E_{j}-E_{j^{\prime}}\right\vert \gg\hbar\omega_{0}$. We conclude that in the
framework of Mesoscopic QED where the cavity frequency pull is a central
quantity, it is particularly important to eliminate in a rigorous way the
$\hat{A}^{2}$ terms if one wants to use a linear light/matter coupling. The
photonic pseudo-potential scheme completes such a task.

The cavity frequency pull is not the only means to characterize the
interactions between a nanocircuit and a cavity. A more general workout of
Mesoscopic QED devices requires the knowledge of the matrix elements between
the different nanocircuit eigenstates, generated by the cavity photons.
Another argument in strong favor of the use of the photonic pseudo-potential
scheme is the general relation
\begin{align}
\left\langle \varphi_{j}\right\vert V_{\perp}(\overrightarrow{\hat{q}_{\alpha
}})\left\vert \varphi_{j^{\prime}}\right\rangle  &  =\frac{ie}{2m_{\alpha
}(E_{j^{\prime}}-E_{j})}\nonumber\\
&  \left\langle \varphi_{j}\right\vert \overrightarrow{\hat{p}}_{\alpha
}.\mathcal{\vec{A}}(\overrightarrow{\hat{q}_{\alpha}})+\mathcal{\vec{A}%
}(\overrightarrow{\hat{q}_{\alpha}}).\overrightarrow{\hat{p}}_{\alpha
}\left\vert \varphi_{j^{\prime}}\right\rangle ~\text{.} \label{el}%
\end{align}
It indicates that the single-photon coupling elements decay more quickly with
$(E_{j^{\prime}}-E_{j})$ in the photonic pseudo-potential scheme. In practice
one often uses a truncated space for the nanocircuit electronic states which
is difficult to handle or determine globally. This is what we do for instance
in section IV, where we consider a single orbital quantum dot. Equation
(\ref{el}) shows that in this case it is more accurate to use
Eq.(\ref{Htottild}) to predict the cavity behavior.

Finally, the photonic pseudo-potential scheme involves only photon-independent
superconducting gap terms (see section \ref{supra}). This can be another
significant advantage considering the already rich structure of the
nanocircuit/light coupling in the absence of superconductivity.


\begin{thebibliography}{99}
\bibitem {reservoir}Strictly speaking, this assumption is not valid for non
ergodic reservoirs such as integrable billiards or Anderson localized systems.
Our theory can be applied to these cases provided one keeps the $j$ dependence
of the coupling parameter $\alpha_{oj}$.
\bibitem {geneBCS}The results of section \ref{supra} can be generalized
straighforwardly to d-wave or p-wave superconductivity by using a non-local
gap term in $\Delta(\vec{r},\vec{r}^{\text{ }\prime})e^{\hat{\Phi}(\vec
{r})+\hat{\Phi}(\vec{r}^{\text{ }\prime})}$ in Eq.(\ref{Hsupr}), with a
different structure in spin-space in case of p-wave superconductivity.
\end{thebibliography}
\end{document}